\documentclass[prx,superscriptaddress,twocolumn,longbibliography]{revtex4-2}

\usepackage{amsmath}
\usepackage{amssymb}
\usepackage{amsxtra,color}
\usepackage{booktabs}
\usepackage{latexsym}
\usepackage{dsfont}
\usepackage{array}
\makeatletter
\@ifundefined{selectlanguage}
  {\newcommand{\selectlanguage}[1]{}}
  {\renewcommand{\selectlanguage}[1]{}}
\makeatother

\usepackage{graphicx}
\usepackage{mathtools}
\usepackage{booktabs}

\DeclareUnicodeCharacter{0394}{\ensuremath{\Delta}}
\DeclareUnicodeCharacter{0393}{\ensuremath{\Gamma}}

\graphicspath{ {attoimages/} }   
\usepackage{braket}
\usepackage{tikz}
\usetikzlibrary{arrows}
\usepackage{pgfplots}
\usepgfplotslibrary{groupplots}
\pgfplotsset{compat=1.3}
\usepackage{lipsum}
\usepackage{color}
\usepackage{comment}

\usepackage{float}

\usepackage[T1]{fontenc}
\usepackage{lmodern}
\usepackage[utf8]{inputenc} 
\usepackage{MnSymbol}	

\usepackage[unicode]{hyperref}

\definecolor{MyDarkGreen}{rgb}{0,0.6,0}
\definecolor{MyDarkBlue}{rgb}{0,0,0.8}
\definecolor{MyDarkRed}{rgb}{0.6,0,0.3}

\hypersetup{breaklinks=true, colorlinks=true,plainpages=true, linktocpage=true, linkcolor=MyDarkBlue, citecolor=MyDarkGreen, urlcolor=MyDarkRed, pdfborder={0 0 0},%
pdfauthor={},%
pdfsubject={Research article},
pdftitle={}%
}

\begin{document}

\title{Coherent phonon motions and ordered vacancy compound mediated quantum path interference in Cu-poor CuIn$_{x}$Ga$_{(1-x)}$Se$_2$ (CIGS) with attosecond transient absorption}
\author{Hugo Laurell}
\email{hugolaurell@lbl.gov}
\affiliation{Department of Chemistry, University of California, Berkeley, California, 94720, USA}
\affiliation{Department of Physics, University of California, Berkeley, California, 94720, USA}
\affiliation{Material Sciences Division, Lawrence Berkeley National Laboratory, Berkeley, California 94720, USA}
\affiliation{Department of Physics, Lund University, Box 118, 22100 Lund, Sweden}
\author{Jonah R. Adelman}
\affiliation{Department of Chemistry, University of California, Berkeley, California, 94720, USA}
\author{Elizaveta \surname{Yakovleva}}
\affiliation{Department of Materials Science and Engineering, Solar Cell Technology, Uppsala University, Box 35,751 03 Uppsala, Sweden}
\author{Carl \surname{Hägglund}}
\affiliation{Department of Materials Science and Engineering, Solar Cell Technology, Uppsala University, Box 35,751 03 Uppsala, Sweden}
\author{Kostiantyn \surname{Sopiha}}
\affiliation{Department of Materials Science and Engineering, Solar Cell Technology, Uppsala University, Box 35,751 03 Uppsala, Sweden}
\author{Axel \surname{Stenquist}}
\affiliation{Department of Physics, Lund University, Box 118, 22100 Lund, Sweden}
\author{Han K. D. \surname{Le}}
\affiliation{Department of Chemistry, University of California, Berkeley, California, 94720, USA}
\author{Peidong \surname{Yang}}
\affiliation{Department of Chemistry, University of California, Berkeley, California, 94720, USA}
\affiliation{Material Sciences Division, Lawrence Berkeley National Laboratory, Berkeley, California 94720, USA}
\affiliation{Department of Materials Science and Engineering, University of California, Berkeley, California 94720, USA}
\affiliation{Kavli Energy Nano Science Institute, Berkeley, California 94720, USA}
\author{Marika \surname{Edoff}}
\affiliation{Department of Materials Science and Engineering, Solar Cell Technology, Uppsala University, Box 35,751 03 Uppsala, Sweden}
\author{Stephen R. \surname{Leone}}
\affiliation{Department of Physics, University of California, Berkeley, California, 94720, USA}
\affiliation{Department of Chemistry, University of California, Berkeley, California, 94720, USA}
\affiliation{Chemical Sciences Division, Lawrence Berkeley National Laboratory, Berkeley, California 94720, USA}

\begin{abstract}
\begin{centering}
In this study, coherent phonon motion is observed in bandgap excited CuIn$_{x}$Ga$_{(1-x)}$Se$_2$ (CIGS) utilizing extreme ultraviolet (XUV) attosecond transient absorption spectroscopy across the Se M$_{4,5}$ absorption edge. Two frequencies of coherent phonon motion are resolved, a low frequency mode attributed through Raman measurements to the $A_{1g}$ phonon motion of a Cu-deficient ordered vacancy compound (OVC), while the high frequency mode originates from the $A_{1g}$ phonon motion in the chalcopyrite phase. The two oscillations lead to modulations in the XUV differential absorption $\Delta A(\epsilon,\tau)$ due to energy shifts of the Se M$_{4,5}$ edge, with a minima occuring approximately 1 ps after the band gap excitation. The hot carrier cooling time of holes and electrons are individually obtained and the observed slower cooling of holes is attributed to the higher density of hole states in the valence band. We also observe fast oscillations (18.6(3) fs period) across the Se absorption edge, which are interpreted to originate from quantum path interference between the electronic conduction bands of the chalcopyrite CIGS and OVC phases, opening the possibility towards quantum coherent metrology in photovoltaics on the femtosecond timescale. The complex interplay between the chalcopyrite and OVC phases are revealed in this investigation through both coherent vibrational and electronic motions.
\end{centering}
\end{abstract}
\maketitle

\renewcommand{\thefootnote}{\textdagger}
\footnotetext{These authors contributed equally to this work.}

\section{Introduction}

With the discovery of high-harmonic generation and attosecond light pulses \cite{McPhersonJOSAB1987,FerrayJPB1988}, electron dynamics have been measured in atomic \cite{OttScience2013}, molecular \cite{Nandi2020} and solid-state systems \cite{LiuNP2016}. A key method for the characterization of electron dynamics on their natural time scale is attosecond transient absorption spectroscopy (ATAS), \cite{WirthScience2011,Attar2020,Palo2024,Zuerch2017,Geneaux20212,Drescher2023} which often has excellent sensitivity to hole and electron dynamics in both the valence and conduction bands. The study of carrier and lattice dynamics in semiconductors on attosecond and femtosecond timescales is crucial for understanding the fundamental processes governing their optical and electronic properties.

In perspective of the global effort of transitioning energy production away from fossil fuels, novel photovoltaic platforms can play a crucial role \cite{victoria_2021}. Photovoltaics now offers the cheapest form of energy production in many countries worldwide, and it is forecast to surpass nuclear and wind electricity generation within five years \cite{iea_renewables_2023}. One such platform of special promise is the \textbf{I-III-VI$_2$} chalcopyrite compounds such as Cu(In$_{x}$Ga$_{x-1}$)Se$_2$ (CIGS) \cite{ramanujam_2017}, with a current record photovoltaic efficiency of $23.6\%$ \cite{Keller2024}. The potential of CIGS lies in its high absorption coefficient, allowing for thin 1-2 $\mu$m thick absorber layers, and direct bandgap that can be tuned by changing the [Ga]/([Ga]+[In]) ratio \cite{ramanujam_2017}.

An attractive application for CIGS is in tandem solar cells, where alloying with Ag and higher levels of Ga to achieve a larger bandgap, promote crystallization, and improve band alignment for the top subcell can substantially improve overall light harvesting efficiency \cite{Keller2020,yang_2022}. In parallel, lower‐bandgap CIGS compositions remain of great interest for single-junction cells that could rival the performance of monocrystalline Si. Notably, Ag alloying, while beneficial for tandem designs\cite{Keller2020}, also enhances the OVC formation\cite{Keller2020_2}. This underscores the importance of systematic studies of the (A)CIGS–OVC system to optimize both band‐edge tuning and heterojunction quality. Ultrafast spectroscopy provides a means to study key parameters in the optimization of solar cell efficiency, such as recombination times, recombination pathways and hot carrier cooling, inaccessible with more traditional phenomenological studies. Detailed characterization of the coherent phonon motion in CIGS could reveal how to efficiently store hot carrier energy in the phonon motion and subsequently transfer it to the carriers (hot-phonon bottleneck). This potentially paves the way toward phonon engineering approaches to use hot-carrier energy to enhance photovoltaic efficiency beyond the Schockley-Queisser limit\cite{chan2021}. 

\begin{figure*}[htp!]
\includegraphics[width=0.9\linewidth]{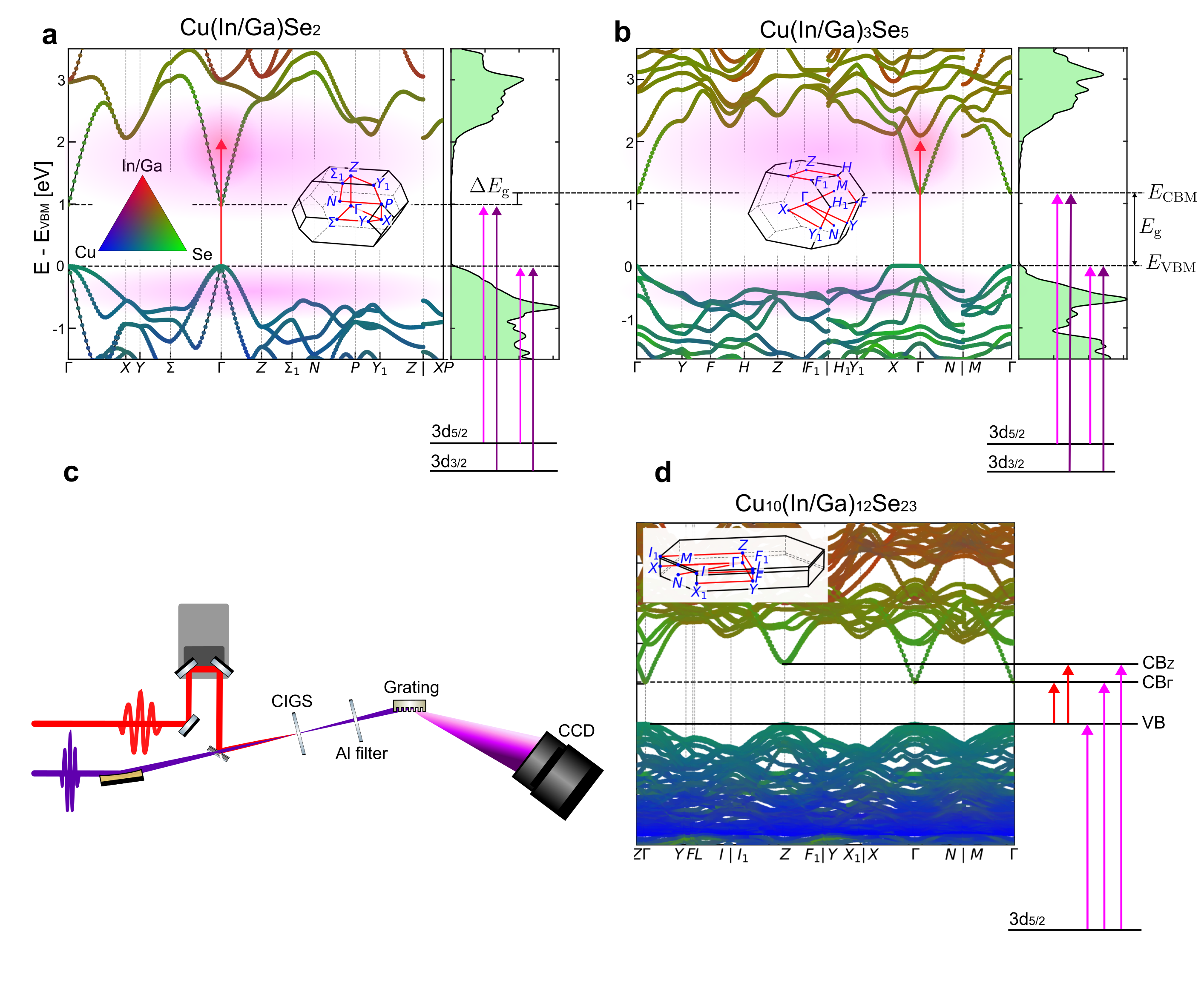}
\caption{In (a,b) the chalcopyrite and OVC band structures calculated using DFT are shown, respectively. Here Cu(In/Ga)Se$_2$ represent the chalcopyrite, while the Cu(In/Ga)$_3$Se$_5$ denotes the OVC. The contributions of Se, In and Cu orbitals are color coded in green, red and blue, respectively. The [Ga]/([Ga]+[In]) ratio was set to 0 for the calculations. The panels on the right show the Se projected density of states. The red arrows indicate the optical excitation. In (c) a schematic of the measurement procedure is shown. A few femtosecond IR pump pulse is sent to the CIGS thin film sample to enable transition across the bandgap. The photoinduced dynamics is then recorded by transmitting an attosecond XUV pulse through the sample and recording the XUV absorption spectrum. By changing the optical path length between the IR pump and XUV probe the effective delay between pump and probe pulses is varied, from which the differential absorption can be reconstructed. In (d) the energy diagram for the 10:12:23 type-II band alignment across the distributed heterojunctions is shown. At pump-probe delay $\tau=0$ an IR photon transfers population from the $\Gamma$ and Z valence bands into the $\Gamma$ and Z conduction bands. The CBs then evolve with different phase due to the conduction band offset $\Delta E_{CB}$. Subsequently, an XUV photon is absorbed from the Se $3d_{5/2,3/2}$ ground state to probe the population in the CBs and VBs of the chalcopyrite and OVC, as well as the coherence between the CBs. The dashed parabolas in the CBs represent the shift in conduction band minimum (CBM) due to the bandgap renormalization, which is modulated by the coherent phonon motion. The red arrows indicate the optical excitations and the purple arrows indicate the XUV probe transitions from the $3d_{5/2}$ core level.} \label{fig1}
\end{figure*}
Core-level XUV transient absorption spectroscopy has enabled the simultaneous observation of carrier and lattice dynamics in semiconductors with unprecedented temporal resolution \cite{GoulielmakisNature2010,Palo2024}. The technique combines both high spectral and temporal resolution, while being especially advantageous for measuring the holes due to their spectral isolation. The core level transitions provide independent probes of the valence and conduction bands, making it ideal to disentangle complex carrier-lattice dynamics. Measurements in Si across the L$_{2,3}$ absorption edge, for example, have recovered the time constants for inter/intra valley scattering and disentangled scattering contributions imparted by electrons and phonons \cite{Cushing2019}. Additionally, studies have been performed on Si nanoparticles to investigate phonon bottleneck effects that slow down the loss of hot carrier energy\cite{Porter2021}. Using femtosecond pump-probe spectroscopy, the dominant recombination mechanism at carrier densities on the order of $10^{20}$ cm$^{-3}$, for co-evaporated CIGS, has been determined to be Shockley-Read-Hall recombination. At these carrier densities the carrier cooling time (equilibration with phonons) has been measured to be $\sim$3 ps \cite{Chen2012}.

Ordered vacancy compounds (OVCs) are believed to play an important role in enhancing the efficiency of the CIGS absorber layer in photovoltaic cells \cite{Zhang1998}. Copper deficiency leads to the formation of an OVC phase adjacent to bulk CIGS, creating buried type-II heterojunctions that optimize charge separation and transport. These heterojunctions create a hole barrier by widening the CIGS band gap through lowering of the valence band maximum (VBM) in the defect phase. This barrier prevents holes from reaching the CIGS/CdS interface, improving charge collection efficiency and reducing recombination losses \cite{Zhao2021,Caballero2010,Schmid1996}. Through this mechanism the OVC enhances the efficiency of CIGS-based solar cells. By understanding the dynamic interplay between the crystal phases, future improvements of CIGS design might be possible, thereby advancing photovoltaic technologies. 

In this work, we have performed ATAS measurements on CIGS thin films across the Se M$_{4,5}$ absorption edge. As the conduction band minimum (CBM) and VBM both have Se character (see Fig. \ref{fig1}), the element specific XUV light probes a spectral region that is central to the photovoltaic performance. Coherent oscillations of the absorption edge are observed as a function of pump-probe delay at energies corresponding to the transition from the $3d_{5/2,3/2}$ core levels to the CBM. Two distinct lower-frequency oscillations are observed at 4.8 and 5.3 THz, corresponding to the $A_{1g}$ phonon modes of the chalcopyrite and the OVC phases, and one higher frequency oscillation is uncovered at 55 THz, originating from quantum path interference between the two phases (electronic coherence). In addition, there is an increase in absorption in the valence band region, indicating the generation of a pump-induced hole population. We employ an iterative decomposition, following the methodology developed by Zuerch \textit{et. al.}\cite{Zuerch2017}, to disentangle the temporal response of holes, electrons and coherent optical phonons.

In Fig. \ref{fig1}(a,b) the band structure of the chalcopyrite phase (a) and OVC (b) are shown, calculated using Density Functional Theory (DFT). The phases are represented by the Ga-free chalcopyrite CuInSe$_2$ ($E_g = 1$ eV) and model CuIn$_3$Se$_5$ structure illustrated below. The calculations were performed using the PBE+U functional, with Hubbard U correction of 5 eV applied on the Cu $3d$ orbitals according to Dudarev \textit{et al.}\cite{Dudarev1998}. The band gaps were scissor-shifted to the experimental values measured for CIGS film herein. The band gap was measured to be $E_g = 1.1$ eV using ellipsometry and is direct at the $\Gamma$-point. The panels on the right in Fig. \ref{fig1}(a,b) show the projected density of states for Se. The valence band predominantly consists of Cu $3d$ and Se $4p$ anti-bonding orbitals, while the conduction band is dominated by Se $4s$ and In $5s$ orbitals. The bandgap difference between the chalcopyrite and OVC is indicated as $\Delta E_g$ and has a value of 0.18 eV \cite{Maeda2016}, while the energy for the CBM and VBM is shown as $E_{CBM}$ and $E_{VBM}$, respectively. The difference in $E_{VBM}$ is due to the fact that the Cu $3d$ and Se $4p$ form anti-bonding orbitals near the VBM, and when Cu atoms are removed to form the OVC this shifts the VBM to lower energies. Experimentally, the Fermi level is above the midgap for the OVC, making it n-type, while the chalcopyrite is strongly p-type\cite{Schmid1993}. Therefore, as the two phases are in contact, p-n heterojunctions are formed.

In Fig. \ref{fig1}(c), a schematic of the ATAS measurements of the thin film CIGS sample is shown. An IR few femtosecond pump pulse optically excites electrons across the bandgap of CIGS, initiating the photoinduced dynamics. Subsequently, the XUV attosecond probe pulse is transmitted through the CIGS film and the transient XUV absorption spectrum recorded by a CCD camera. By recording the XUV absorption spectra, with and without infrared pump, for a set of pump-probe delays the differential absorption, $\Delta A(\epsilon,\tau)$, is reconstructed. In Fig. \ref{fig1}(d), a simplified energy diagram, with respect to the Se 3$d_{5/2}$ core level, of the involved states that partake in the optical transitions at the distributed heterojunctions is shown. The structure is the Cu$_{10}$(In/Ga)$_{12}$Se$_{23}$, which includes both the OVC and the chalcopyrite unit cells in its supercell, more details on this in section \ref{short}. Excitation of an electron from the VB to the CB via the absorption of an IR photon generates hole and electron populations at the $\Gamma$ or Z-point. The localization of carriers to the $\Gamma$ and Z-point is also due to the fact that for direct transitions the bandgap is larger than the IR photon energy at all other points. The momentum mismatch between the $\Gamma$ and Z point is small, and indirect transitions between these valleys could be weakly allowed. Furthermore, the prominence of the Z-valley increases with reduced Cu concentration, implying that it is a signature of the ordered vacancies. The bandgap difference of the two valleys is a reflection of the larger bandgap for the OVC/Z due to the reduced antibonding in the VB. The difference in XUV absorption, with and without IR pump, probe the valence and conduction band populations through state blocking (SB) of the XUV transition. Due to the proximity between Se and Ge in the periodic table, and the similar core transitions, we anticipate the core hole to be well screened in Se, as in Ge \cite{Zuerch2017}.

\section{Methods}
\subsection{Sample synthesis}
For CIGS deposition, the Si\(_3\)N\(_4\) substrates were placed in a titanium holder with an opening just below the size of 5 mm x 5 mm. For the purpose of optical characterization, one 5 mm x 15 mm Si substrate piece was also included in the run. To prevent diffusion of copper into the Si during CIGS deposition, a barrier layer of approximately 20 nm thick hafnium oxide was first grown on this Si substrate using atomic layer deposition (ALD) at 250°C, with tetrakis(dimethylamido)hafnium (Sigma-Aldrich, 99.99\%) and water as precursors.

CIGS was deposited in vacuum through co-evaporation from fixed open boat sources with evaporation rates controlled by mass-spectrometer feedback. All elements were evaporated at constant rates during the entire deposition time, aiming for a slightly Cu-poor stoichiometry with a Cu/(Ga+In) ratio of 0.9 and a [Ga]/([Ga]+[In]) ratio of 0.26. The ratios of elements were confirmed with X-ray fluorescence (XRF) measurement on a sample with \(1 \; \mu m\) CIGS thickness. The deposition time was then scaled down in the next deposition run to result in a sample of 50 nm thick CIGS.

The samples were heated from the back by infrared heaters, and they reached a maximum of 425°C during the deposition process. The deposition time window started with opening the shutter in front of the samples when stable temperature and evaporation rates had been achieved, and the deposition lasted for 75 s.
After CIGS deposition and cooling down in vacuum, the samples were unloaded from the vacuum system and transferred within 5 min to an ALD system. Using trimethyl aluminum (Pegasus, electronic grade) and water, a layer of approximately 10 nm thick of Al\(_2\)O\(_3\) was grown at 120°C on all samples to suppress CIGS oxidation.

Spectroscopic ellipsometry measurements were performed on the Si substrate before and after the Al\(_2\)O\(_3\) growth to determine the thickness and bandgap of the CIGS layer. A generalized oscillator model including one Psemi-m0 and 9 Gaussians was employed to represent the CIGS layer. The model also included a top layer based on the Bruggeman effective medium approximation, to take CIGS roughness, with and without the Al\(_2\)O\(_3\), into account. The resulting CIGS thickness was slightly less than 48 nm. The extrapolated bandgap was 1.1 eV, which corresponds well to the expected [Ga]/([Ga]+[In])$ ~=0.26$ ratio \cite{Boyle2014, Keller2020}.

\subsection{Core-level XUV transient absorption measurements}
The XUV transient absorption spectroscopy measurements of the CIGS samples were performed by photo-exciting CIGS with a few fs nominally 800 nm IR pump pulse and waiting a delay $\tau$ until a probe XUV pulse was transmitted through the sample, and the XUV absorption spectrum recorded. The beam path starts with 35 fs, 2.3 mJ pulses with a central wavelength of 800 nm (IR) at 1 kHz repetition rate delivered from a Ti:Sapphire Coherent Astrella laser system. The pulses are then sent through a stretched 2 m long hollow core fiber filled with neon for spectral broadening. The spectrally broadened laser pulses were then temporally compressed to 4 fs using chirped mirrors and a piece of z-cut Ammonium Dihydrogen Phosphate (ADP) to compensate for $3^{rd}$ order dispersion accumulated in the fiber. Subsequently, the IR pulses were sent into a Mach-Zehnder interferometer, where in the probe arm the IR pulse was focused with a mirror with $f = 50$ cm into a gas cell filled with argon. In the focus, few-burst attosecond pulses are generated through the process of high-order harmonic generation (HHG). The residual IR was filtered out with an aluminum foil leaving only the XUV pulse. Then, the XUV was recombined with the pump IR pulses on an annular mirror and both the IR and XUV were focused on the CIGS sample. The IR pump excites an estimated carrier density of $ 2\cdot10^{20}$ cm$^{-3}$ in the CIGS layer. After transmission through the sample, a secondary aluminum foil was used to filter out the residual pump IR. Finally, the transmitted XUV is diffracted by a grating onto a CCD (PIXIS 400 B) where it was recorded.

\section{Results and Discussion}
In this work, we present four sets of measurements, where the pump-probe delay was varied with different time steps and over different time delay ranges to capture slow and fast time scales of the photo-induced dynamics. Long scans ($\sim$ps) were recorded to study the coherent phonon motion and hot carrier cooling in CIGS, while short scans ($\sim$100 fs) were performed to study fast electron dynamics in CIGS across the Se M$_{4,5}$ absorption edge.

\begin{figure}[h]
\includegraphics[width=\linewidth]{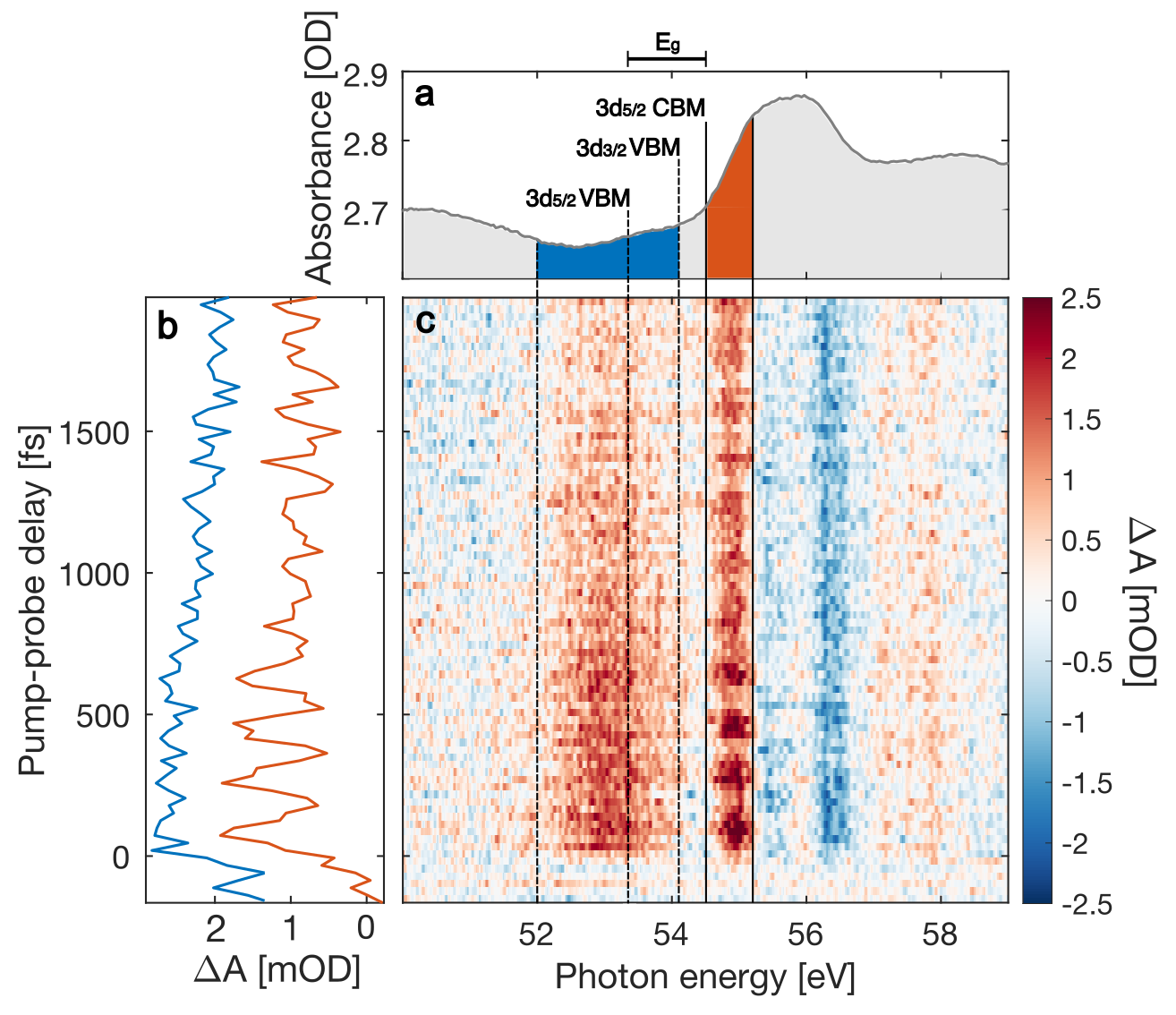}
\caption{XUV absorption spectrum (a), spectral averages of the VB and CB spectral regions (b) and transient absorption scan of CIGS (c). The delay was varied in steps of $26.4$ fs between $-165.0$ and $1973.4$ fs. The VBM probed via the $3d_{3/2}$ transition was determined from the zero-crossing of the temporally averaged $\Delta A(\epsilon,\tau)$ and was found to $E_\text{VBM} = 54.1$ eV. The VB spectral region ((a) blue region) was determined as the interval from the low energy zero crossing of $\Delta A$ at the $E_\text{VBM}$. By subtracting the spin-orbit splitting energy from $E_\text{VBM}$ and adding the bandgap energy, the CBM probed via the $3d_{5/2}$ transition was found at $E_\text{CBM} = 54.5$ eV. The upper bound of the CB region ((a) red region) probed via the $3d_{5/2}$ transition was determined from the zero crossing of the $\Delta A(\epsilon,\tau)$ to 55.2 eV. To the spectral average of VB (in (b)), a 1 mOD offset has been added for visibility.\label{fig2}}
\end{figure}

In Fig. \ref{fig2} (c), an attosecond transient absorption scan of the CIGS sample is shown. The delay was scanned between -165 and 1973.4 fs with steps of 26.4 fs. The colormap shows the differential absorption ($\Delta A(\epsilon,\tau)$) in units of mOD, where a positive $\Delta A(\epsilon,\tau)$ indicates an increase in absorption and vice versa. In the top panel (Fig. \ref{fig2} (a)), the XUV absorption spectrum is shown. The VBM probed via the $3d_{3/2}$ transition was determined from the zero crossing of $\Delta A(\epsilon,\tau)$ to $E_\text{VBM} = 54.1$ eV, agreeing with photoelectron spectroscopy measurements \cite{Nelson1997}, while the lower bound of the VB spectral region was determined to 52 eV (Fig. \ref{fig2} (a) blue region). Subsequently, the CBM probed via the $3d_{5/2}$ transition was determined to be $E_\text{CBM} = 54.5$ eV, by subtracting the Se $3d$ spin-orbit splitting energy from the $3d_{3/2}$ $E_\text{VBM}$ and adding the band gap energy. The upper bound on the CB spectral region (Fig. \ref{fig2} (a) red region) was determined via the zero crossing of the $\Delta A(\epsilon,\tau)$ to 55.2 eV. 

The spectral region from 52.0 to 54.1 eV (blue shaded in the XUV absorption spectrum) shows an increase in absorption, this we attribute to the opening of transitions to the VB from the Se $3d$ shell. The second region of interest (red shaded in the XUV absorption spectrum in Fig. \ref{fig2}(a)) shows an oscillation in the $\Delta A$ as a function of pump-probe delay in Fig. \ref{fig2}(c). The differential absorption here corresponds predominantly to XUV transitions from the $3d_{5/2}$ ground state to the conduction band. The oscillation in $\Delta A$ as a function of delay is typical of coherent phonon motion \cite{Geneaux2021} and the results are consistent with the frequencies of the two $A_{1g}$ phonon modes measured for the chalcopyrite and the OVC phase with Raman spectroscopy (see SM \ref{ramann}). To the best of our knowledge this is the first observation of coherent phonon dynamics in CIGS. The two coherent phonons are likely generated by the change in potential (force) from depopulation of Cu $3d$ and Se $4p$ anti-bonding orbitals in the VB of CIGS \cite{Ghosh2002}.

In the CB spectral region we have three main contributions to the $\Delta A(\epsilon,\tau)$: (1) Redshift of the absorption edge due to bandgap renormalization (BGR) and a change in core-level binding energy, (2) state-blocking due to the IR mediated population transfer from the VB to the CB, labeled carriers, as well as (3) excited state broadening \cite{Zuerch2017}. Due to the 0.65 eV spin-orbit splitting energy of Se $3d$ core levels, which has been measured in CIGS\cite{Nelson1997}, there is partial spectral overlap between the state blocking and lattice features that correspond to the core-level transitions from the $3d_{3/2,5/2}$ shells. The left panel (Fig. \ref{fig2} (b)) shows the spectral average of differential absorption across the valence band and conduction band regions. For the holes there is a fast increase in absorption around $t=0$ followed by a slow decay over several ps. For the CB there is a fast increase in absorption followed by the onset of an oscillation in $\Delta A(\epsilon,\tau)$. Importantly, there is a dephasing of the beats in the oscillation with a minimum around $1$ ps and a subsequent revival.

\subsection{Iterative decomposition of the differential absorption}

The variation of $\Delta A(\epsilon,\tau)$ within the spectral region around 55 eV originating from the transition from the Se $3d_{5/2}$ ground state to the CB has several contributions listed in the previous section. To disentangle the individual components, an iterative decomposition algorithm was implemented, following the methodology developed by Zuerch \textit{et al.} \cite{Zuerch2017}, decomposing the differential absorption according to,
\begin{equation}\label{uno}
\begin{split}
\Delta A(\epsilon,\tau) &= \Delta A_\text{shift}(\epsilon,\tau) + \Delta A_\text{broadening}(\epsilon,\tau) \\&+ \Delta A_\text{carriers}(\epsilon,\tau).
\end{split}
\end{equation}
\begin{figure*}[htp!]
\includegraphics[width=\linewidth]{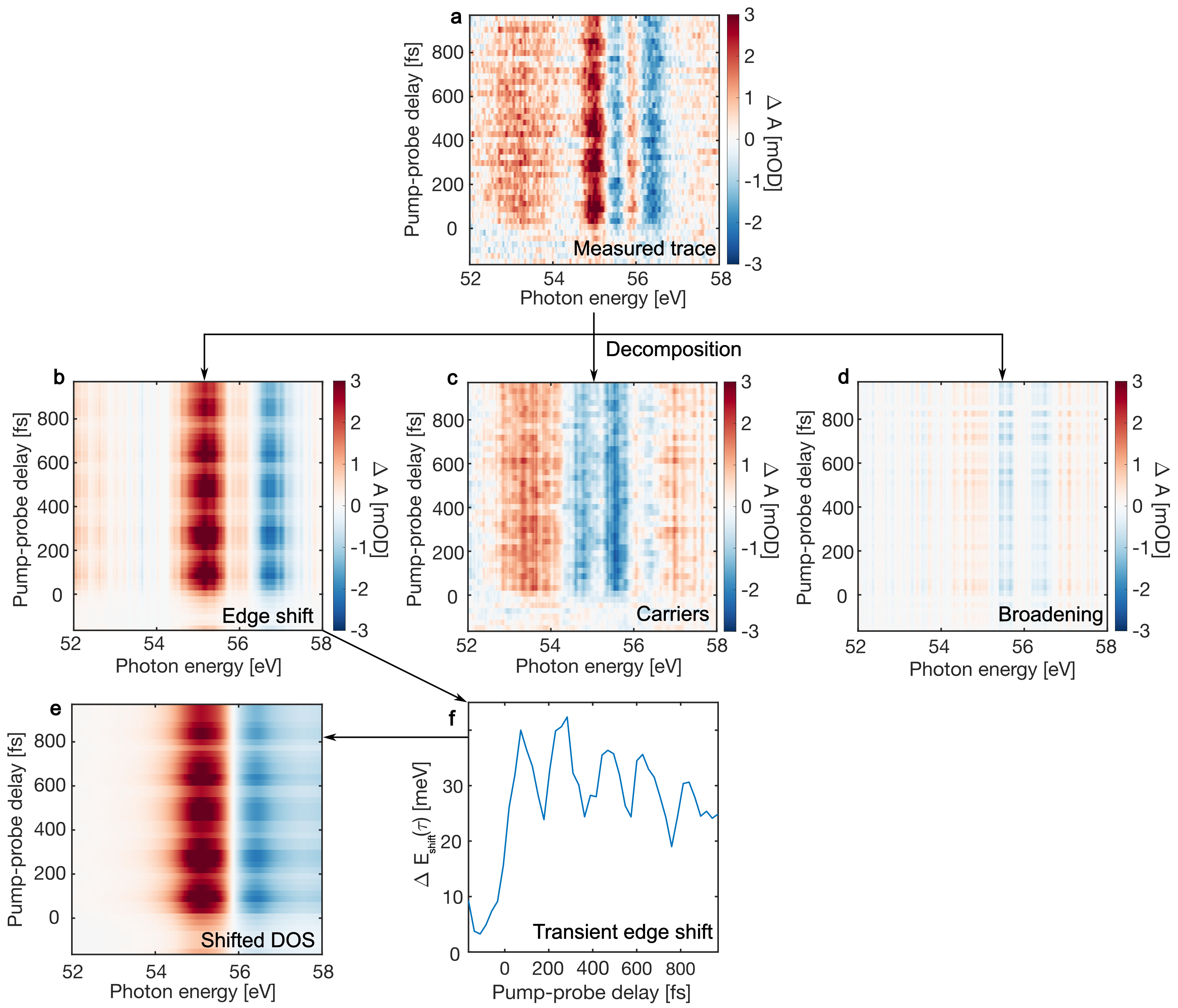}
\caption{Iterative decomposition of transient absorption scan. (a) Raw transient absorption scan. The raw scan is separated into the underlying spectral contributions that include absorption edge shift (b), state blocking by carriers (c), and excited state broadening (d). The two negative state blocking features in (c) are due to spin-orbit replicas from the spin-orbit split Se core level. (e) $\Delta A_\text{shift}(\epsilon,\tau)$ is computed from the total CB Se DOS for the chalcopyrite and OVC using the retrieved transient edge shift $\Delta E(\tau)$ and Eq. \eqref{hatti}. In (f) the retrieved transient edge shift, $\Delta E_\text{shift}(\tau)$ is shown.\label{decomp}}
\end{figure*}
In Fig. \ref{decomp} the iterative decomposition of the measured 1 ps scan (a) is shown. The contributions due to shifting of the absorption edge (b), State blocking by carriers (c) and Excited state broadening (d) are indicated. The excited state broadening does not vary significantly as a function of pump-probe delay over 1 ps. Moreover the excited state broadening in Fig. \ref{decomp} (d) is at the noise level and its contribution to the $\Delta A$ can therefore be neglected. The coherent phonon dynamics is well described by the shifting of the absorption edge, which is retrieved from the decomposition (see Fig. \ref{decomp} (b)). The transient edge shift originates from a modulation of the $E_{CBM}$ and Se $3d$ core level binding energies. Following a similar procedure as was made in the case of silicon \cite{Cushing2018}, we calculate the initial shift in core level binding energy as $45$ meV. In addition, since the core level shift counteracts the BGR, we can estimate the magnitude of the inital BGR as $11$ meV. For a transition from the Se $3d$ ground state the edge shift contribution to the differential absorption can be written as,
\begin{equation}\label{hatti}
\begin{split}
\Delta A_\text{shift}(\epsilon,\tau) &= \sum_j \alpha_j(A_j(\epsilon + \Delta E_\text{shift}(\tau)) - A_j(\epsilon)).
\end{split}
\end{equation}
Where $A_j(\epsilon)$ is the XUV absorption spectra of the $j$-th crystal phase and $\Delta E_\text{shift}(\tau)$ is the transient edge shift. The fractions of contributions for the two phases ($\alpha_j$) were extracted from the Raman measurements (see Fig. \ref{Raman} in the SM). The core hole broadening is described with a convolution of the Lorentzian lineshape function, $L(\epsilon)$, with a FWHM of $\Gamma = 0.8$ eV \cite{Bahl1980}. In Fig. \ref{decomp} (b) the edge shift contribution to the differential absorption is shown, $\Delta A_\text{shift}(\epsilon,\tau)$, assuming identical edge shifts for the chalcopyrite and OVC phases. From the transient edge shift $\Delta E_\text{shift}(\tau)$ (Fig. \ref{decomp} (f)), the DFT calculated Se-projected density of states and the Lorentzian core hole broadening, the theoretical differential absorption can be calculated according to Eq. \eqref{hatti}, as shown in Fig. \ref{decomp} (e). There is good agreement between the theoretical differential absorption and $\Delta A_\text{shift}(\epsilon,\tau)$, i.e. Fig. \ref{decomp}(b) and (e).

\subsection{Temporal analysis of carriers and phonons}
To characterize the photo-induced dynamics in CIGS, the decay of the electron and hole populations ($\tau_\text{electron}^\text{decay}$, $\tau_\text{hole}^\text{decay}$), the common decoherence time of the coherent phonons ($T_2^{ph}$) and phonon phases ($\phi^\text{CIGS}_{A_{1g}},\phi^\text{OVC}_{A_{1g}}$) were fitted for the spectral averages of the VB and CB regions indicated in Fig. \ref{fig2}. The rise and decay of the differential absorption ($\Delta A(\epsilon,\tau)$) were fitted using exponentially modified Gaussian functions, which take into account the effect of the cross-correlation between the IR pump and XUV probe pulses.
\begin{figure}[h]
\includegraphics[width=\linewidth]{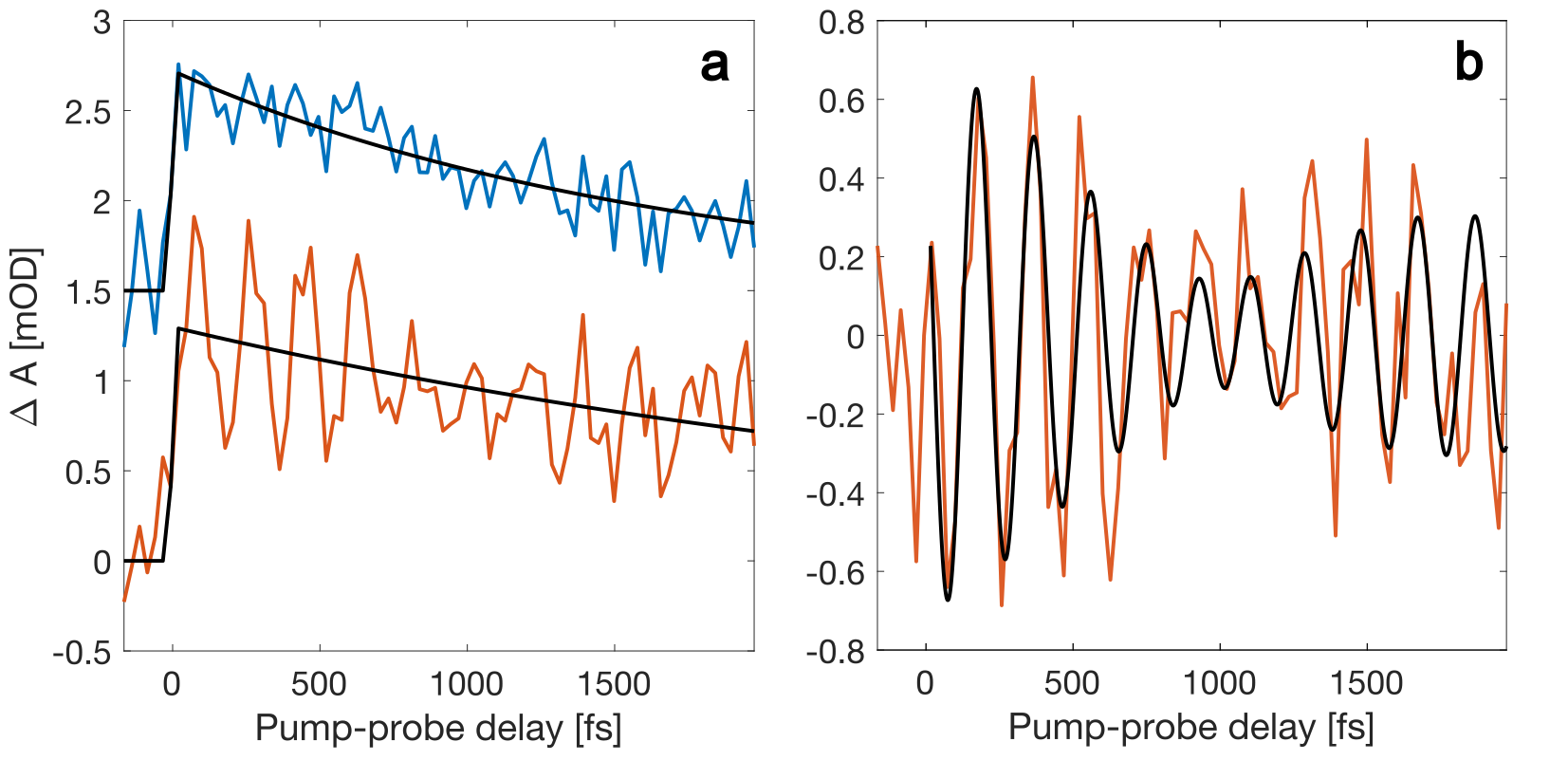}
\caption{In (a) the spectral average of $\Delta A(\epsilon,\tau)$ shows the differential absorption as a function of pump-probe delay time for the valence (blue) and conduction bands (red), respectively. The black curves represent the fit of the rise and decay using an exponentially modified Gaussian function. In (b) the residual oscillating component ($\Delta A_\text{phonon}(\tau)$) of the conduction band in (a) is plotted after subtraction of the fitted rise and decay. \label{fig:3}}
\end{figure}
In Fig. \ref{fig:3} (a) the spectral averages of the two spectral regions indicated in Fig. \ref{fig2} corresponding to the VB and CB response are shown. At $t=0$ there is a sharp increase in absorption for both curves followed by an exponential decay. Superimposed on the relaxation of the lattice (red curve) there are clear oscillations with a period of $\sim 200$ fs. At around $t = 1$ ps there is a minimum in the oscillation amplitude, followed by a revival, indicating that there is a dephasing between the two phonon modes due to their different frequencies. The rise and decay of the two spectral averages were fitted, where the fit results are plotted as black curves in Fig. \ref{fig:3} (a). The relaxation of the lattice (based on the edge shift) and holes were fitted to, $\tau^{\text{decay}}_\text{phonon} = 2.7(8)$ ps, $\tau^{\text{decay}}_\text{hole} = 1.7(2)$ ps, where the lattice relaxation was fitted from the decomposed edge shift contribution ($\Delta E_\text{shift}(\epsilon,\tau)$) in Eq. \eqref{uno} of the 1 ps scan. The 1.7 ps decay of the holes is similar to the previously reported carrier cooling time of $\sim 3$ ps for co-evaporated CIGS at a carrier density of $2\cdot10^{20}$ cm$^{-3}$ \cite{Chen2012}. The mechanism for the carrier cooling has previously been determined to be electron-phonon scattering \cite{Othonos1998}. For reference, the characteristic time for recombination of photo-carriers is two orders of magnitude slower, e.g. $\sim$ 750 ps recombination decay time was measured in co-evaporated CIGS with similar carrier concentrations. The vastly different timescale means the carrier concatenations cannot be resolved from our measurements. In addition, the dominant recombination mechanism was determined to be Shockley-Read-Hall for co-evaporated CIGS \cite{Chen2012}. The decay of the CB was fitted from $\Delta A_\text{carriers}(\epsilon,\tau)$, retrieved from the decomposition, to $\tau_\text{electron}^\text{decay} = 2.6(5)$ ps. Bringing everything together, we conclude that the hole cooling time is approximately half that of the cooling time of the electrons and the relaxation of the lattice. The larger DOS in the VB compared to the CB is a likely origin for the higher scattering rate for the holes as compared to the electrons. Specifically, since the electron-phonon scattering matrix elements incorporate the electron DOS, more available states for holes to scatter into gives a higher rate and shorter cooling time. 

To characterize the coherent phonon motion, the fit of the rise and decay in the CB spectral region were subtracted from the spectral average, shown in Fig. \ref{fig:3} (b), leaving the oscillating coherent phonon component of the differential absorption. To probe the phonon frequencies, Raman measurements (shown in Fig. \ref{Raman} in the Supplemental Material) were performed on the CIGS samples and two peaks at $\nu^\text{OVC}_{A_{1g}}=159(4)$ cm$^{-1}$ and $\nu^\text{CIGS}_{A_{1g}}=176.2(7)$ cm$^{-1}$ were observed. These are identified in previous work and are assigned as the $A_{1g}$ coherent phonon of the chalcopyrite phase and its spectrally shifted copy in the ordered vacancy compound \cite{SHEU2016}. The relative amplitudes from Raman spectroscopy were obtained as $\alpha^\text{CIGS}_{A_{1g}} = 65\%$ and $\alpha^\text{OVC}_{A_{1g}} = 35\%$. The frequencies $(\omega_{A_{1g}}^\text{CIGS},\omega_{A_{1g}}^\text{OVC})$ and relative amplitudes $(\alpha_{A_{1g}}^\text{CIGS},\alpha_{A_{1g}}^\text{OVC})$ retrieved from the Raman measurements, were incorporated into a fit of the coherent phonons to reduce the number of parameters of the fit function. The results show that there will be dephasing and recurrences due to the difference in phonon frequencies, as well as a permanent loss of coherence, or decoherence. In Fig. \ref{fig:3} (b) the fit is plotted as a solid black curve, where the fit function used was parameterized as,
\begin{equation}
\begin{split}
\Delta A_\text{phonon}(\tau)  & = Ce^{-\frac{\tau}{T_2^{ph}}}\bigg(\alpha_{A_{1g}}^\text{OVC}\cos(\omega_{A_{1g}}^\text{OVC}\tau + \phi_{A_{1g}}^\text{OVC})\\& + \alpha_{A_{1g}}^\text{CIGS}\cos(\omega_{A_{1g}}^\text{CIGS}\tau + \phi_{A_{1g}}^\text{CIGS})\bigg).\label{residual}
\end{split}
\end{equation}
Here $C$ is the oscillation amplitude, $T_2^{ph}$ is the coherent phonon decoherence time and ($\phi_{A_{1g}}^\text{CIGS}$,$\phi_{A_{1g}}^\text{OVC}$) are the phonon phases for chalcopyrite and OVC, respectively. It is assumed that the two phonons decohere on the same timescale $T_2^{ph}$, a reasonable assumption as the chalcopyrite and OVC lattices are very similar. The fitted phonon phases were, $\phi^\text{OVC}_{A_{1g}} = 0.73(4)$ rad and $\phi^\text{CIGS}_{A_{1g}} = 0.75(2)$ rad, with the shared decoherence time of $T_2^{ph} = 2.27(8)$ ps. The decoherence mechanism of the coherent phonons is likely through Klemens decay where one coherent optical phonon is converted into two acoustic phonons \cite{Klemens1966}. Since the phonon phases, $(\phi^\text{CIGS}_{A_{1g}},\phi^\text{OVC}_{A_{1g}})$, are similar and close to $\pi/4$, the mechanism responsible for the excitation of the coherent phonon is likely a mixture of displacive excitation of coherent phonons (DECP) and impulsive stimulated Raman scattering  (ISRS)\cite{Zeiger1992}. For ISRS the phonon phase is predicted to be $\pi/2$, while for DECP the value should be closer to 0 or $\pi$. In the DECP model, the driving force of the coherent phonon motion can originate from either population transfer $n(t)$ or variations in electronic temperature $\Delta T_e(t)$ \cite{Zeiger1992,Geneaux2021,Drescher2023,Adelman2025}. The most likely driving force in CIGS is population transfer as CIGS lacks a Peierls distortion\cite{Drescher2023}.

\subsection{Few femtosecond time-scale dynamics}\label{short}
\begin{figure*}[htbp!]
\includegraphics[width=\linewidth]{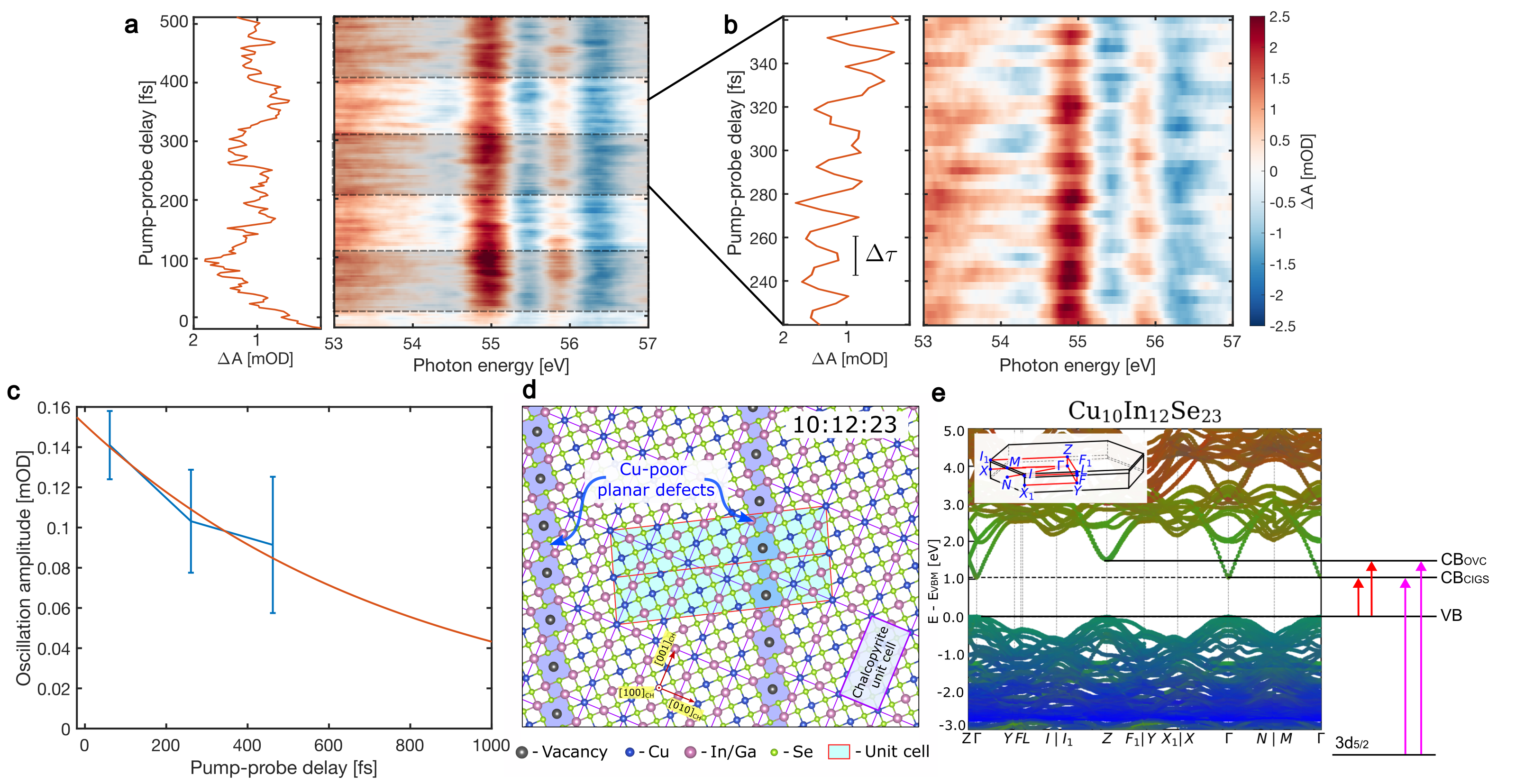}
\caption{Attosecond transient absorption scans of CIGS where the delay is scanned from $-19.8$ to $511.5$~fs (a) and $219.8$ to $361.7$~fs (b) with a time step of $3.3$~fs. In the left panels in (a,b), the spectral averages over the CB regions are plotted, showing clear oscillations in $\Delta A$. In the left panel in (b) $\Delta \tau$ indicates the fitted period of oscillation of $18.6(3)$~fs. In (c) the fitted oscillation amplitude of the bandpass filtered lineout of the CB region in (a) is shown over the first three phonon maxima. From the decay of the oscillation amplitude, the fit of the decoherence time is $0.8(4)$~ps. In (d) the crystal structure of the chalcopyrite and OVC mixture is shown, showing the Cu poor planar defects separated by 3 nm. At these interfaces the distributed heterojunctions are formed. In (e) the energy diagram for the 4-level model is shown.\label{fig:55}}
\end{figure*}
To investigate further, given the high temporal resolution of ATAS, we conducted measurements in CIGS with a $3.3$~fs time step across the three first phonon oscillations to investigate ultrafast carrier dynamics. In Fig. \ref{fig:55} ATAS delay scans of CIGS are shown where the delay is varied from $-19.8$ to $511.5$~fs in (a) and, focusing on a narrower delay range, from $219.8$ to $361.7$~fs in (b) in steps of 3.3 fs. In the left panels in Fig. \ref{fig:55}(a,b), the spectral averages over the CB region reveal clear oscillations of the absorption edge with a fitted period of $\Delta \tau = 18.6(3)$~fs across the second phonon oscillation maximum. These rapid oscillations are interpreted as a two-optical-path interference between the OVC and chalcopyrite CB’s, owing to the type-II band alignment \cite{Zhang1998}. However, this explanation requires significant overlap between wavefunctions of both phases, which can only occur if the interface-to-bulk ratio is large. This situation is possible in CIGS because of the inherent off-stoichiometry, mediated by the formation of thermodynamically stable OVC-like inclusions inside the otherwise ideal CIGS grains, as illustrated in Fig. \ref{fig:55}(d) \cite{Sophia2022}. These inclusions are Cu-poor and have local bonding structure similar to that in the Cu(In,Ga)$_3$Se$_5$, but they cannot be strictly characterized as an OVC phase due to the small thickness. Judging by the composition of our films (Cu/(Ga+In)=0.9), the average distance between the OVC planar defects is estimated to be $\sim$ 3 nm. Provided the OVC-like inclusions are electronically similar to the OVC phase, which usually exhibits n-type conductivity \cite{Zhang1998}, the bulk of the CIGS film constitutes a distributed heterojunction. Following the optical excitation, at $\tau = 0$, photo-exited electrons are generated in the state of a coherent superposition. This coherence is traced in time by probing optical transitions from the Se 3d shell into the CB using attosecond XUV pulses.

In Fig. \ref{fig:55} (e) the DFT calculated band structure of the 10:12:23 supercell is shown, which includes both the calcopyrite and OVC unit cells. In contrast to the chalcopyrite and OVC band structures in Fig. \ref{fig1}(a,b), a new CB at the Z point emerges due to the inclusion of the OVC. Since the VBM in the chalcopyrite and the CBM in the OVC lie at the $\Gamma$ and Z point respectively, the optical transitions between these manifolds are not momentum-conserving, however provided that the initial and final state wavefunctions sufficiently overlap they could be weakly allowed. This facilitates direct interfacial charge transfer, a phenomenon predicted and experimentally observed in other material platforms. Examples include momentum-matched type-II van der Waals heterostructures \cite{Zhang2023}, direct photoluminescence-based evidence of type-II transitions in a ZnO/P3HT junction \cite{Chan2012}, and observations of both direct type-I and interface-related indirect type-II transitions in InGaAs/AlAsSb quantum wells \cite{Mozume2000}. Additionally, studies of WS$_2$/Bi$_2$O$_2$Se heterojunctions have demonstrated programmable interfacial band alignment, enabling type-II direct transitions and localized interlayer excitons dependent on $\Gamma$-point wavefunction overlap  \cite{Zhang2024}. Further support of this interpretation comes from observed strong photoresponses attributed explicitly to interfacial transitions in type-II junctions, with clear threshold energies indicating genuine interfacial transition mechanisms \cite{Cao2012}. These studies collectively provide credence to interpreting the rapid oscillations as a consequence of direct interfacial charge transfer. Nevertheless, the strength and feasibility of such direct transitions critically rely upon the symmetry match of wavefunctions at the interface, an important aspect requiring detailed theoretical studies beyond the scope of this work.

Within this interpretation, the measured period corresponds to a conduction band minima offset of, $h/\Delta\tau=\Delta E_{CB} = 0.222(7)~\text{eV}$. This value is in line with theoretical studies of the conduction band offset between the chalcopyrite phase and OVC across distributed heterojunctions \cite{Zhang1998}, and it agrees with the extrapolated value from DFT calculations $\Delta E^{th}_{CB} = 0.25$ eV of the CIGS sample in this study (Fig. \ref{fig:dCBM} in the SM). To model the photo-induced dynamics, we consider direct optical transitions from the chalcopyrite valence band to the OVC and chalcopyrite conduction bands in a simple 4-level model (shown in Fig. \ref{fig:55}(e)). The 4-level model consists of the states, $\ket{VB}$, the two CBs $\ket{CB_{1}}$ and $\ket{CB_2}$, where the VB is the chalcopyrite VB and the indices $\{1,2\}$ correspond to OVC and the chalcopyrite, respectively. The core level is represented by $\ket{3d_{5/2}}$. Solving the Lindblad master equation perturbatively to first order (full derivation shown in SM \ref{cohcoh}) yields an evolving coherence between the CB states as,
\begin{equation}\label{fnaff}
\begin{split}
\rho^{(1)}_{CB_1,CB_2}(\tau) &= \mu_{VB,CB_1}\mu_{VB,CB_2}^*\rho_{CB_1,CB_2}(0) \\
&\quad\times \exp\!\Biggl(-\frac{i}{\hbar}\Delta E_{CB}\,\tau - \Gamma\,\tau\Biggr),
\end{split}
\end{equation}
where $\Delta E_{CB} = E_{CB_2} - E_{CB_1}$ is the energy difference between the conduction band minima of the two crystal phases and $\Gamma$ is the decoherence rate. The XUV mediated transition from the core level then probes the coherence between the conduction bands, resulting in the state blocking contribution to the differential absorption signal of the form (full derivation in SM \ref{cohcoh})
\begin{equation}\label{fnaff2}
\begin{split}
\Delta A_\text{carriers}(E_{f},\tau) &\propto -\rho_{CB_1,CB_1}(0)-\rho_{CB_2,CB_2}(0) \\&- 2|\rho_{CB_1,CB_2}(0)|\,e^{-\Gamma \tau}\cos\!\Biggl(\frac{\Delta E_{CB}\,\tau}{\hbar}+\phi\Biggr),
\end{split}
\end{equation}
where, $\phi = \mathrm{Arg}(\rho_{CB_1,CB_2}(0))$, is the phase of the initial coherence. One can note that the expression of the state blocking contribution in Eq. \eqref{fnaff2} is always negative, as state blocking cannot increase the XUV transition amplitude to the CB's. Notably, we do not observe oscillations in the VB region. This is due to the fact that no coherent superposition is created in the chalcopyrite VB, as it has a single degree of crystal phase freedom. By fitting the oscillation amplitude in Fig. \ref{fig:55}(a) across the three first phonon oscillations ($\sim$100 fs intervals outlined by the shaded regions in Fig. \ref{fig:55}(a)), the blue points in Fig. \ref{fig:55}(c) are produced with associated uncertainty. Fitting the decay of the oscillation amplitude, weighted with the inverse of the error from the fit of the oscillation amplitude, the decoherence time $1/\Gamma = T_{2}^e$ is extracted as 0.8(4) ps. Additionally, we fit the phase over the first 100 fs and relate the imprinted phase shift in the differential absorption to a time delay experienced for the electron while crossing the junction. We find a phase shift of \(\phi = 4.1(3)\)~rad, which corresponds to a time delay of \(t_W = 12.1(8)\)~fs. Assuming that contributions from the dipole transition matrix elements (both from the IR and core-level transitions) are negligible, this time delay can be interpreted as the time it takes for the electron to traverse the junction. 

Furthermore, we can try to estimate the degree of coherence between the conduction bands from the scan in Fig. \ref{fig:55}(b), recognizing the limitations of the estimate. From the Raman measurements we can approximate the populations of the two conduction bands, $(\rho_{CB_1,CB_1},\rho_{CB_2,CB_2})$, as the normalized Raman intensities for the two phonon modes. We can also infer the coherence $|\rho_{CB_1,CB_2}|$ from the oscillation amplitude of the rapid oscillations, full details in SM \ref{cohish}. From the estimated elements of the density matrix we retrieve the degree of coherence between the chalcopyrite and OVC conduction bands at the second phonon oscillation as,
\begin{equation}
 C = \frac{|\rho_{CB_1,CB_2}|}{\sqrt{\rho_{CB_1,CB_1}\rho_{CB_2,CB_2}}} \approx 0.19.
\end{equation}
The partial coherence of roughly $20\%$ is due to a number of decoherence channels. Present decoherence channels include experimental decoherence, in the form of temporal jitter between the pump and probe and camera noise, as well as decoherence due to interactions with the environment, such as electron-phonon and electron-electron scattering. We also know that intra-phase transitions, originating from spatial regions where no junctions are present, will contribute to a reduced degree of coherence as pure state blocking. If only spatial regions with wavefunction overlap between the chalcopyrite and OVC were probed, the observed degree of coherence would likely be higher.

We hypothesize that the observed long-lived electronic coherence ($T_2^e = 0.8(4),\mathrm{ps}$), exceeding typical semiconductor coherence times (e.g., $12.9,\mathrm{fs}$ in GaAs at room temperature \cite{Takagi2023}), originates from the the spatial separation of the OVC and chalcopyrite conduction bands and the coherent phonon motion. The spatial separation of the conduction band states across the buried type-II chalcopyrite/OVC heterojunctions \cite{Sato2014} leads to reduced wavefunction overlap, suppressing interphase electron-electron and electron-phonon scattering. As a consequence, the populations remain approximately time-invariant during the measurement, consistent with the assumptions for the perturbative solutions to the Lindblad equation (Eq. \eqref{fnaff2}). Within this framework, this leaves intraphase scattering as the primary dephasing mechanism. In intraphase scattering within each CB manifold, electrons can scatter internally with phonons or other electrons. Intraphase electron-phonon scattering, in particular, is closely related to carrier cooling, for which we measure a cooling time of the electrons to $2.6(5),\text{ps}$, significantly longer than the observed coherence time. However, cooling involves multiple sequential phonon emission events, each imparting incremental and possibly random phase shifts. Since the conduction band minima (CBM) for the chalcopyrite and OVC phases lie at the $\Gamma$ and Z-point respectively, intraphase scattering primarily involves zone center phonons. The $A_{1g}$ phonons are coherent during our measurement and electron scattering on these coherent phonons could possibly explain the origin of the extended coherence time. Other possible explanations for the long lived electronic coherence include correlated lattice fluctuations as observered in perovskite quantum dots \cite{Ghosh2025}, or as a result of coherence between incoherent excitons \cite{Zhang2025}. Additionally, intraphase electron-electron scattering will contribute to the dephasing and if dominant would lead to a rapid dephasing, disagreeing with our measurements. Therefore, we suggest that the long electronic coherence time results from a combination of suppressed interphase scattering due to spatial localization of electrons in the distinct crystal phases and intraphase scattering via coherent optical phonons at the distributed chalcopyrite/OVC heterojunctions. 

To test this hypothesis, future experiments could investigate the dependence of the electronic coherence time on carrier density. Such studies could probe the relative contributions of interphase and intraphase scattering mechanisms and determine the extent to which spatial localization at the distributed heterojunctions influence ultrafast electronic decoherence. To summarize, the measurements provide compelling evidence for efficient, ultrafast interfacial charge transfer enabled by wavefunction overlap at the chalcopyrite/OVC interface, highlighting the critical role of atomic-scale interface design in optimizing photovoltaic performance.

\section{Summary}
In this study, we present the first attosecond transient absorption spectroscopy (ATAS) characterization of thin-film Copper Indium Gallium Selenide (CIGS) samples. Utilizing high temporal and spectral resolution, we observe the coherent phonon motion corresponding to the $A_{1g}$ modes of both the chalcopyrite phase and a Cu-deficient ordered vacancy compound (OVC) phase, with frequencies of $\nu^\text{CIGS}_{A_{1g}}=176.2(7)$ cm$^{-1}$ and $\nu^\text{OVC}_{A_{1g}}=159(4)$ cm$^{-1}$, respectively. The superposition of the two phonon oscillations results in a beating in differential absorption with a minimum of the oscillation amplitude around 1 ps in the differential absorption, $\Delta A(\epsilon,\tau)$, and later recurrence. Additionally, we measure rapid oscillations with a fitted period of 18.6 fs across the Se M$_{4,5}$ absorption edge, which we attribute to quantum path interference between the chalcopyrite and OVC phase. The contrast of the high frequency oscillations encodes the degree of electronic coherence, which could be a potential new observable for photovoltaic performance. Through an iterative decomposition of the transient absorption data, we successfully disentangled the contributions from electron and hole cooling as well as the lattice response, providing insights into the carrier cooling and coherent phonon motion in CIGS. The measured hot hole cooling time of approximately 1.7 ps and hot electron cooling time of 2.7 ps aligns with previous studies and improves the understanding of carrier thermalization in CIGS. These processes are inseparable parts of photo excitations in CIGS-based solar cells, which on the fundamental level are likely to underpin their exceptional photovoltaic performance. The ability to disentangle coherent phonon motion and hot carrier cooling at femtosecond to picosecond timescales opens new avenues for tailoring material properties to achieve higher solar cell efficiencies. This work underscores the potential of ATAS as a powerful tool for probing ultrafast dynamics in complex semiconductor systems, paving the way for future advancements in photovoltaic technology. 

\section{Acknowledgements}
H.L would like to thank J. M. Dahlström for fruitful discussions. This work was supported by the Department of Energy, Office of Science, Basic Energy Science (BES) Program within the Materials Science and Engineering Division (contract DE-AC02-05CH11231), through the Fundamentals of Semiconductor Nanowires Program through the Lawrence Berkeley National Laboratory. Support for laser instrumentation and vacuum hardware is from AFOSR grant numbers FA9550-19-1-0314, FA9550-24-1-0184, and FA9550-22-1-0451. H.L. acknowledges support from the Swedish Research Council (2023-06502) and the Sweden-America Foundation. J.R.A acknowledges the NSF GRFP under grant No DGE 2146752. H.K.D.L acknowledges support from the National Science Foundation’s Graduate Research Fellowship Program (NSF GRFP) under grant DGE 1752814

\bibliography{Ref_lib}

\clearpage
\onecolumngrid

\section{Supplemental material}

\subsection{Data analysis}

The analysis of the differential absorption scans were performed as follows, first the energy axis was calibrated to the absorption of the singly excited states in Ne, $2s^22p^6 ~^1S_0\rightarrow 2s2p^6np ~^1P_0$, $n = \{3,4,5\}$, at well known energies of 45.547, 47.123 and 47.694 eV \cite{Codling1967}. Then, the scan was re-binned using a cubic spline interpolation on the linear energy grid calibrated from the Ne XUV absorption measurement. Subsequently, an edge referencing algorithm was applied \cite{Geneaux2021} to reduce noise induced by high-harmonic spectral drifts. The edge referencing regions are chosen as $[38.0,45.1]$ eV and $[59.2,70.0]$ eV, as in these spectral regions there were no measurable differential absorptions above the noise floor. In the scans to investigate the femtosecond dynamics, with delay step 3.3 fs, temporal drifts between pump and probe caused an increased noise level; this necessitated a shorter scan time, in turn leading to a lower SNR. Therefore a moving average was applied to these scans to increase visibility of the spectral features. The rapid oscillations ($\sim$55 THz) were reproduced in three separate measurements, the two presented in the manuscript and a third where the delay was varied in steps of 2.9 fs. All uncertainties reported in this manuscript are retrieved from the $1\sigma$ confidence intervals of the fits.

\subsection{Solution of the Lindblad master equation for the 4-level model}\label{cohcoh}

We consider a four-level system with the following states, a ground state \( \ket{VB} \), two excited states $\ket{CB_1}$ and $\ket{CB_2}$ with energies $E_{CB_1}$ and $E_{CB_2}$ and an energy splitting $\Delta E_{CB} = E_{CB_2}-E_{CB_1}$, and a core level $\ket{3d_{5/2}}$ with energy $E_{3d_{5/2}}$. The energies are ordered as $E_{3d_{5/2}} < E_{VB} < E_{CB_1} < E_{CB_2}$. Here $\ket{CB_1}$ represent the OVC conduction band while $\ket{CB_2}$ the CIGS conduction band and $\ket{VB}$ the CIGS valence band. An impulsive IR pump pulse at \( \tau=0 \) creates a coherent superposition of \( \ket{CB_1} \) and \( \ket{CB_2} \). After a delay \( \tau \), the XUV probe pulse excites the core level electron \( \ket{3d_{5/2}} \) to the conduction band states. We assume the probe is weak and impulsive so that its effect can be treated perturbatively. The observable is the differential absorption \( \Delta A_\text{SB}(E,\tau) \) as a function of XUV photon energy \( E_f \) and pump-probe delay \( \tau \). By selecting energies \( E_f \approx E_{CB_j} - E_{3d_{5/2}} \), we probe the coherence of the excited state. In what follows we derive the differential absorption $\Delta A_\text{SB}(E,\tau)$, including pure dephasing via the Lindblad master equation, and show that the differential absorption is proportional to the imaginary part of the Fourier transform of the induced macroscopic polarization similar to published work \cite{Pfeiffer2012}. Initially the system is in the ground state,
\begin{equation}
\rho(t<0) = \ket{VB}\bra{VB}\,.
\end{equation}
Where $\rho$ is the density matrix. The IR pump pulse (at \( t=0 \)) excites the system into the superposition state
\begin{equation}
\ket{\psi(0)} = a\,\ket{CB_1} + b\,\ket{CB_2}\,,
\end{equation}
with complex amplitudes \(a\) and \(b\). In density matrix representation the IR prepared system reads as,
\begin{equation}
\begin{split}
\rho_{CB_1,CB_1}(0) &= |a|^2 = \frac{\alpha_1}{\alpha_1 + \alpha_2} = 0.65\\
\rho_{CB_2,CB_2}(0) &= |b|^2 = \frac{\alpha_2}{\alpha_1 + \alpha_2} = 0.35\\
\rho_{CB_1,CB_2}(0) &= a\,b^* \\
\rho_{CB_2,CB_1}(0) &= a^*\,b\\
\end{split}
\end{equation}
Where $\alpha_j$ is the concentration of the crystal phases, which is known from the Raman measurements. Between pump and probe (\(0 < t < \tau\)) the system evolves freely under \(H_0\). Since \(H_0\ket{CB_j} = E_{CB_j}\ket{CB_j}\), the state \( \ket{CB_j} \) evolves as
\begin{equation}
\ket{CB_j(t)} = e^{-iE_{CB_j}t/\hbar}\ket{CB_j}\,.
\end{equation}
Thus, the off-diagonal element of the density matrix evolves as
\begin{equation}
\rho_{CB_1,CB_2}(t) = \rho_{CB_1,CB_2}(0)\, \exp\!\Bigl[-\frac{i}{\hbar}(E_{CB_1}-E_{CB_2})t\Bigr]\,.
\end{equation}
At \(t=\tau\) this becomes
\begin{equation}
\rho_{CB_1,CB_2}(\tau) = a\,b^*\,\exp\!\Bigl[-\frac{i}{\hbar}(E_{CB_1}-E_{CB_2})\tau\Bigr]\,.
\end{equation}
This phase evolution represents the quantum beat between \( \ket{CB_1} \) and \( \ket{CB_2} \). At \(t=\tau\) the XUV probe pulse interacts with the system. The probe couples the core state \( \ket{3d_{5/2}} \) to the conduction band states via the dipole operator
\begin{equation}
\mu = \mu_{3d_{5/2},CB_1}\ket{3d_{5/2}}\bra{CB_1} + \mu_{3d_{5/2},CB_2}\ket{3d_{5/2}}\bra{CB_2} + \text{h.c.}
\end{equation}
We assume that the core level, $3d_{5/2}$, is populated before the interaction with the probe, so that \(\rho_{3d_{5/2},3d_{5/2}}(\tau)\) is nonzero. The probe is treated as a weak, impulsive perturbation, with its effect given by
\begin{equation}
\rho^{(1)}(\tau) = -\frac{i}{\hbar}\Bigl[ V_{XUV},\,\rho(\tau) \Bigr]\,,
\end{equation}
where the effective XUV interaction is
\begin{equation}
V_{XUV} = -\Bigl[\mu_{3d_{5/2},CB_1}\ket{3d_{5/2}}\bra{CB_1} + \mu_{3d_{5/2},CB_2}\ket{3d_{5/2}}\bra{CB_2} + \text{h.c.}\Bigr]\,.
\end{equation}
We focus on the density matrix element
\begin{equation}
\rho_{3d_{5/2},CB_1}^{(1)}(\tau) = \bra{3d_{5/2}}\rho^{(1)}(\tau)\ket{CB_1}\,.
\end{equation}
Expanding the commutator gives
\begin{equation}
\rho_{3d_{5/2},CB_1}^{(1)}(\tau) = -\frac{i}{\hbar}\Bigl( \bra{3d_{5/2}}V_{XUV}\,\rho(\tau)\ket{CB_1} - \bra{3d_{5/2}}\rho(\tau)\,V_{XUV}\ket{CB_1} \Bigr)\,.
\end{equation}
The first term,
\begin{equation}
T_1 = \bra{3d_{5/2}}V_{XUV}\,\rho(\tau)\ket{CB_1}
\end{equation}
receives a contribution from the component of \(V_{XUV}\) that connects \(\ket{CB_1}\) and $\ket{CB_2}$ to \(\ket{3d_{5/2}}\). We have,
\begin{equation}
\begin{split}
T_1 &= -\mu_{3d_{5/2},CB_1}\,\langle{3d_{5/2}}\ket{3d_{5/2}}\,\bra{CB_1}\rho(\tau)\ket{CB_1}-\mu_{3d_{5/2},CB_2}\,\langle{3d_{5/2}}\ket{3d_{5/2}}\,\bra{CB_2}\rho(\tau)\ket{CB_1} \\
&= -\mu_{3d_{5/2},CB_1}\,\rho_{CB_1,CB_1}(\tau)-\mu_{3d_{5/2},CB_2}\,\rho_{CB_2,CB_1}(\tau)\,.
\end{split}
\end{equation} 
The second term,
\begin{equation}
T_2 = \bra{3d_{5/2}}\rho(\tau)\,V_{XUV}\ket{CB_1},
\end{equation}
receives a contribution from the component of \(V_{XUV}\) that connects \(\ket{3d_{5/2}}\) to \(\ket{CB_1}\). Thus,
\begin{equation}
T_2 = -\mu_{3d_{5/2},CB_1}\,\bra{3d_{5/2}}\rho(\tau)\ket{3d_{5/2}} = -\mu_{3d_{5/2},CB_1}\,\rho_{3d_{5/2},3d_{5/2}}(\tau)\,.
\end{equation}
Hence, the induced coherence is
\begin{equation}
\rho_{3d_{5/2},CB_1}^{(1)}(\tau) = \frac{i}{\hbar}\Bigl[\mu_{3d_{5/2},CB_2}\,\rho_{CB_2,CB_1}(\tau)+\mu_{3d_{5/2},CB_1}\,\rho_{CB_1,CB_1}(\tau)-\mu_{3d_{5/2},CB_1}\,\rho_{3d_{5/2},3d_{5/2}}(\tau)\Bigr]\,.
\end{equation}
A similar expression can be obtained for \(\rho_{3d_{5/2},CB_2}^{(1)}(\tau)\) as, 
\[
\rho^{(1)}_{3d_{5/2}, CB_2}(\tau) = \frac{i}{\hbar}\left[
\mu_{3d_{5/2}, CB_1}\,\rho_{CB_1, CB_2}(\tau)+\mu_{3d_{5/2}, CB_2}\,\rho_{CB_2, CB_2}(\tau)
- \mu_{3d_{5/2}, CB_2}\,\rho_{3d_{5/2}, 3d_{5/2}}(\tau)
\right]\,.
\]
For \(t>\tau\) the induced coherences oscillate at their corresponding frequencies. For example,
\begin{equation}
\rho_{3d_{5/2},CB_1}(t) = \rho_{3d_{5/2},CB_1}(\tau)\,\exp\!\Bigl[-\frac{i}{\hbar}(E_{3d_{5/2}}-E_{CB_1})(t-\tau)\Bigr]\,.
\end{equation}
The macroscopic polarization is given by
\begin{equation}
P(t)=\text{tr}[\mu\,\rho(t)]\,,
\end{equation}
Since we are performing core level spectroscopy, we will select the channels with energy differences corresponding to the core-excited coherences. Thus,
\begin{equation}
P(t) \sim \mu_{3d_{5/2},CB_1}\,\rho_{CB_1,3d_{5/2}}(t) + \mu_{3d_{5/2},CB_2}\,\rho_{CB_2,3d_{5/2}}(t) + \text{c.c.}
\end{equation}
For \(t>\tau\) these coherences oscillate at the frequencies,
\begin{equation}
\omega_{3d_{5/2},CB_1}=\frac{E_{3d_{5/2}}-E_{CB_1}}{\hbar} \quad\text{and}\quad \omega_{3d_{5/2},CB_2}=\frac{E_{3d_{5/2}}-E_{CB_2}}{\hbar}\,.
\end{equation}
Taking the Fourier transform of \(P(t)\) with respect to time gives the polarization as,
\begin{equation}
P(E_f,\tau) = \int_{-\infty}^{\infty} dt\, e^{i\omega_f t}P(t,\tau)\,,
\end{equation}
with \(E_f=\hbar\omega_f\). In the impulsive limit this becomes,
\begin{equation}
\begin{split}
P(E_f,\tau) &\propto P(E_{CB_1}-E_{3d_{5/2}},\tau)\,\delta\!\Bigl(E_f-[E_{CB_1}-E_{3d_{5/2}}]\Bigr) \\
&+ P(E_{CB_2}-E_{3d_{5/2}},\tau)\,\delta\!\Bigl(E_f-[E_{CB_2}-E_{3d_{5/2}}]\Bigr)\,,
\end{split}
\end{equation}
where the Fourier amplitudes of the macroscopic polarization are given by,
\begin{equation}
\begin{split}
P(E_{CB_1}-E_{3d_{5/2}},\tau) &\propto - \frac{i}{\hbar}\mu_{3d_{5/2},CB_1}\mu_{CB_1,3d_{5/2}}\rho_{3d_{5/2},3d_{5/2}}(\tau) \\
&+ \frac{i}{\hbar}\mu_{3d_{5/2},CB_1}\mu_{CB_2,3d_{5/2}}\rho_{CB_2,CB_1}(0)\,e^{-i\Delta E_{CB}\tau/\hbar}\\\
&+\frac{i}{\hbar}\mu_{3d_{5/2},CB_1}\mu_{CB_1,3d_{5/2}}\rho_{CB_1,CB_1}(0)\\
P(E_{CB_2}-E_{3d_{5/2}},\tau) &\propto -\frac{i}{\hbar}\mu_{3d_{5/2},CB_2}\mu_{CB_2,3d_{5/2}}\rho_{3d_{5/2},3d_{5/2}}(\tau) \\
&+\frac{i}{\hbar}\mu_{3d_{5/2},CB_2}\mu_{CB_1,3d_{5/2}}\rho_{CB_1,CB_2}(0)\,e^{i\Delta E_{CB}\tau/\hbar}\\
&+\frac{i}{\hbar}\mu_{3d_{5/2},CB_2}\mu_{CB_2,3d_{5/2}}\rho_{CB_2,CB_2}(0)
\end{split}
\end{equation}
For core level XUV spectroscopy the transmitted XUV field \(E_{\text{out}}(\omega,\tau)\) is related to the incident field \(E_{\text{in}}(\omega)\) and the induced polarization \(P(\omega,\tau)\), assuming that the Beer-Labert law is valid, via
\begin{equation}
E_{\text{out}}(\omega,\tau) = E_{\text{in}}(\omega)\text{exp}\left[i\frac{2\pi\omega}{c}\frac{P(\omega,\tau)L}{E_\text{in}(\omega)}\right]
\end{equation}
The state blocking contribution to the absorption is then given as,
\begin{equation}
\begin{split}
A_\text{SB}(\omega,\tau) &= \log\left(\frac{|E_\text{in}(\omega)|^2}{|E_\text{out}(\omega,\tau)|^2}\right) \\
&=-i\frac{2\pi\omega}{c}\frac{P(\omega,\tau)L}{E_\text{in}(\omega)} + i\frac{2\pi\omega}{c}\frac{P^*(\omega,\tau)L}{E_\text{in}(\omega)}\\
&=2\text{Re}\left(-i\frac{2\pi\omega}{c}\frac{P(\omega,\tau)L}{E_\text{in}(\omega)} \right)\\
&\propto \text{Re}\left(-iP(\omega,\tau)\right) = \text{Im}\left(P(\omega,\tau)\right)
\end{split}
\end{equation}
where \(L\) is the sample thickness. Therefore, in the weak-probe limit the absorption is directly proportional to the imaginary part of the Fourier transform of the macroscopic polarization. In our ideal resolution case, at energies \(E\approx E_{CB_1}-E_{3d_{5/2}}\) and \(E\approx E_{CB_2}-E_{3d_{5/2}}\) the differential absorption becomes,
\begin{equation}\label{fhuefh}
\begin{split}
\Delta A_\text{SB}(E_f,\tau) &= A_\text{SB}^\text{pump off}(E_f,\tau) - A_\text{SB}(E_f,\tau)\\
&\propto \Delta A_\text{SB}(E_{CB_1}-E_{3d_{5/2}},\tau)\,\delta\!\Bigl(E_f-[E_{CB_1}-E_{3d_{5/2}}]\Bigr) \\
&+ \Delta A_\text{SB}(E_{CB_2}-E_{3d_{5/2}},\tau)\,\delta\!\Bigl(E_f-[E_{CB_2}-E_{3d_{5/2}}]\Bigr)
\end{split}
\end{equation}
Where $A^\text{pump off}_{SB}$ is the XUV absorption without IR excitation across the bandgap. Hence in this case $\rho_{CB_1,CB_1} = \rho_{CB_2,CB_2}$ = $\rho_{CB_1,CB_2}$ =  0. Inserting the perturbative expansion of the density matrix into the Eq. \eqref{fhuefh} gives,
\begin{equation}
\begin{split}
\Delta A_\text{SB}(E_{CB_1}-E_{3d_{5/2}},\tau) &\propto \text{Re}\bigg(|\mu_{3d_{5/2},CB_1}|^2\rho_{3d_{5/2},3d_{5/2}}(\tau) \\
&- \mu_{3d_{5/2},CB_1}\mu_{3d_{5/2},CB_2}^*\rho_{CB_2,CB_1}(0)\,e^{-i(E_{CB_2}-E_{CB_1})\tau/\hbar}\\
&-|\mu_{3d_{5/2},CB_1}|^2\rho_{CB_1,CB_1}(0)-|\mu_{3d_{5/2},CB_1}|^2\rho_{3d_{5/2},3d_{5/2}}(\tau)\bigg)\\
&=-|\mu_{3d_{5/2},CB_1}|^2\rho_{CB_1,CB_1}(0) \\
&- |\mu_{3d_{5/2},CB_1}||\mu_{3d_{5/2},CB_2}||\rho_{CB_2,CB_1}(0)|\cos\left(\frac{\Delta E_{CB} \tau}{\hbar} + \phi\right).\\
\end{split}
\end{equation}
That is, the differential absorption measured at the transition \(E=E_{CB_1}-E_{3d_{5/2}}\) contains a static contribution (from the population in \(CB_1\)) and an oscillatory contribution (from the coherence \(\rho_{CB_2,CB_1}\)), the latter yielding quantum beat oscillations with frequency
\begin{equation}
\Delta\omega= \frac{E_{CB_2}-E_{CB_1}}{\hbar} = \frac{\Delta E_{CB}}{\hbar}\,.
\end{equation}
Replacing the discrete CB states with bands of states ($\ket{CB_j} \rightarrow \int dE_f f_j(E_f)\ket{j,E_f}$) with spectral overlap, at a selected energy bin there will be two contributions to the differential absorption, one for the CB of OVC and one for the CB of CIGS. Where $f_j(E_f)$ is the amplitude of the states. Giving the total expression for the differential absorption as,
\begin{equation}
\begin{split}
\Delta A_\text{SB}(E_f,\tau)& \propto  -|\mu_{3d_{5/2},CB_1}|^2\rho_{CB_1,CB_1}(0) 
- |\mu_{3d_{5/2},CB_2}|^2\rho_{CB_2,CB_2}(0)\\
&- 2|\mu_{3d_{5/2},CB_2}||\mu_{3d_{5/2},CB_1}||\rho_{CB_1,CB_2}(0)|\cos\bigg(\frac{\Delta E_{CB}\tau}{\hbar} + \phi\bigg)
\end{split}
\end{equation}
Assuming similar dipole couplings between the states gives,
\begin{equation}
\begin{split}
\Delta A_\text{SB}(E_f,\tau)& \propto -\rho_{CB_1,CB_1}(0) - \rho_{CB_2,CB_2}(0) - 2|\rho_{CB_1,CB_2}(0)|\cos\bigg(\frac{\Delta E_{CB}\tau}{\hbar} + \phi\bigg).
\end{split}
\end{equation}
From this expression we see that the state blocking contributes with a negative differential absorption, which agrees with the observations. Under the approximation of similar strengths of dipole coupling the phase becomes,
\begin{equation}
\phi = \text{Arg}(\rho_{CB_1,CB_2}(0))
\end{equation}
With associated scattering time delay,
\begin{equation}
\tau_W = \hbar \frac{\partial \phi}{\partial E} \approx \hbar\frac{\phi}{\Delta E_{CB}}.
\end{equation}
In an open quantum system, decoherence causes the off-diagonal coherence \( \rho_{CB_1,CB_2} \) to decay. The inclusion of decoherence is facilitated by adding terms to the Liouville von-Neumann equation and instead solving the Lindblad master equation,
\begin{equation}
\frac{d\rho}{dt} = -\frac{i}{\hbar}[H_0,\rho] + \sum_{j=1}^{2}\left( L_j\,\rho\,L_j^\dagger - \frac{1}{2}\{L_j^\dagger L_j,\rho\} \right)\,,
\end{equation}
with collapse operators defined as,
\begin{equation}
L_j = \sqrt{\gamma_j}\,\ket{CB_j}\bra{CB_j}\quad (j=1,2)\,,
\end{equation}
where \( \gamma_j \) are the dephasing rates. For pure dephasing the populations remain unchanged but the off-diagonal element decays between pump and probe $(0<t<\tau)$ according to,
\begin{equation}\label{knork}
\frac{d}{dt}\rho_{CB_1,CB_2}(t) = -\frac{i}{\hbar}(E_{CB_1}-E_{CB_2})\,\rho_{CB_1,CB_2}(t) - \frac{1}{2}(\gamma_1+\gamma_2)\,\rho_{CB_1,CB_2}(t)\,.
\end{equation}
The solution to Eq. \eqref{knork} is on the form,
\begin{equation}\label{fnirk}
\rho_{CB_1,CB_2}(t) = \rho_{CB_1,CB_2}(0)\,\exp\!\Biggl[-\frac{i}{\hbar}(E_{CB_1}-E_{CB_2})t - \frac{1}{2}(\gamma_1+\gamma_2)t\Biggr]\,.
\end{equation}
By defining the average dephasing rate as follow,
\begin{equation}
\Gamma = \frac{\gamma_1+\gamma_2}{2}\,,
\end{equation}
we obtain a new form of Eq. \eqref{fnirk} as,
\begin{equation}
\rho_{CB_1,CB_2}(t) = \rho_{CB_1,CB_2}(0)\,\exp\!\Bigl[-\frac{i}{\hbar}(E_{CB_1}-E_{CB_2})t - \Gamma\,t\Bigr]\,.
\end{equation}
This shows that at \(t=\tau\) the coherence will decay with the factor \(e^{-\Gamma \tau}\). Following a similar calculation as above, the final expression of the differential absorption is given as,
\begin{equation}
\begin{split}
\Delta A_\text{SB}(E_f,\tau)& \propto -\rho_{CB_1,CB_1}(0) - \rho_{CB_2,CB_2}(0) - 2|\rho_{CB_1,CB_2}(0)|e^{-\Gamma \tau}\cos\bigg(\frac{\Delta E_{CB}\tau}{\hbar} + \phi\bigg).
\end{split}
\end{equation}
We have derived the energy- and delay-resolved transient absorption signal \( \Delta A_\text{SB}(E,\tau) \) for a four-level system under impulsive pump and XUV probe excitation. By taking the Fourier transform of the induced macroscopic polarization, we showed that the differential absorption is proportional to the imaginary part of \(P(E,\tau)/E_{\text{in}}(\omega)\). In the derivation, selecting \(E_f \approx E_{CB_1}-E_{3d_{5/2}}\) or \(E_f \approx E_{CB_2}-E_{3d_{5/2}}\) isolates the contributions from the corresponding core-excited transitions. The signal exhibits quantum beat oscillations with frequency \(\Delta \omega=(E_{CB_2}-E_{CB_1})/\hbar\) due to the coherent superposition prepared by the pump. Including dephasing via the Lindblad equation introduces an exponential damping \(e^{-\Gamma \tau}\) of the oscillations. This complete treatment shows how both the electronic coherence and dephasing is reflected in the observed transient absorption.

\subsection{Estimate of degree of coherence}\label{cohish}

From the iterative decomposition of the 1 ps scan we observe that the relative contribution to the total differential absorption of the edge shift is $\Delta A_\text{shift} \approx 5$ mOD and that the relative carrier contribution was $\Delta A_\text{carriers} \approx -2$ mOD, with a total $\Delta A$ of 3 mOD. This gives a relative carrier to lattice ratio of $\Delta A_\text{carriers}/\Delta A_\text{shift} \approx -2/5$. In the scan in Fig. \ref{fig:55}(a) we observe a total $\Delta A$ of 2 mOD at the first phonon maximum. Assuming that the relative carrier contribution of the $\Delta A$ is conserved between the scans, this obtains the carrier contribution at the first phonon maximum, as $\Delta A_\text{carriers} \approx -4/3$. This implies that to relate the fitted oscillation amplitude to the off diagonal element of the density matrix we should normalize as $|\rho_{1,2}| \approx A_\text{osc}/(2|\Delta A_\text{carriers}|) = 0.09$. Where $A_\text{osc}$ is the fitted oscillation amplitude across the second phonon maxima. Furthermore we can estimate the populations of the density matrix as the normalized amplitudes from the Raman measurements, This gives a population for OVC as $\rho_{1,1} = 0.35$ and for CIGS, $\rho_{2,2} = 0.65$. From the estimated elements of the density matrix we calculate the degree of coherence between the CIGS and OVC conduction bands as,
\begin{equation}
C = \frac{|\rho_{1,2}|}{\sqrt{\rho_{1,1}\rho_{2,2}}} \approx 0.19.
\end{equation}

\subsection{Raman Measurements}\label{ramann}

\begin{figure}[h]
    \centering
    \includegraphics[width=0.5\linewidth]{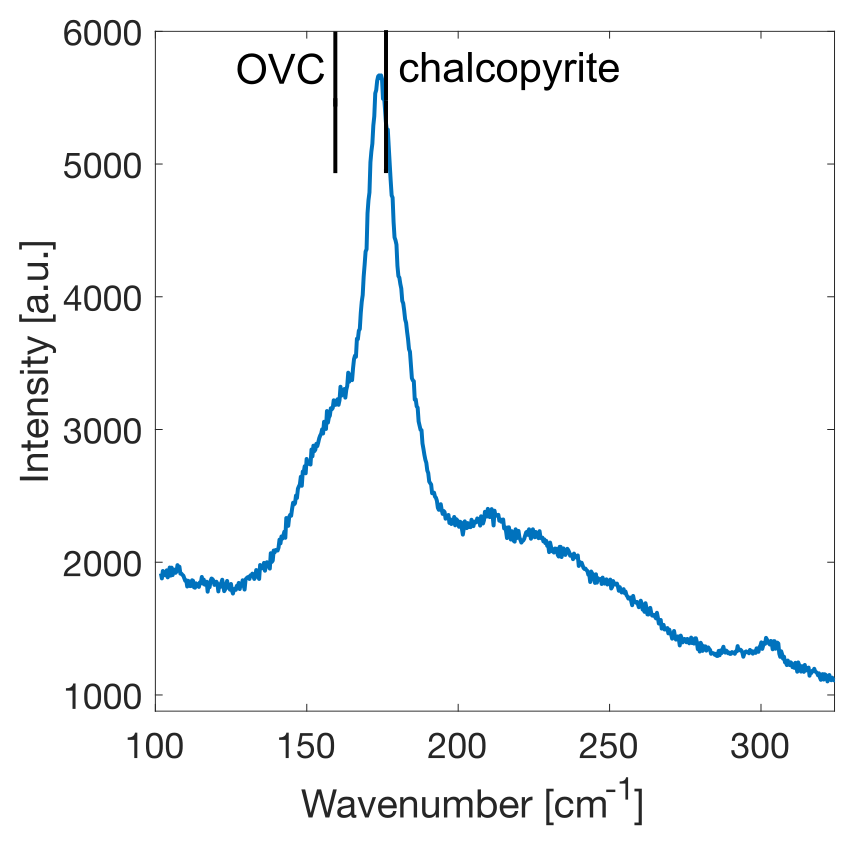}
    \caption{Raman spectra of the CuIn$_{x}$Ga$_{(1-x)}$Se$_2$ sample used in the ATAS experiments. Two features are evident, a shoulder centered at 159(4)~cm$^{-1}$ and a narrow peak at 176.2(7)~cm$^{-1}$. The black lines show the fitted wavenumbers for OVC and the chalcopyrite.}
    \label{Raman}
\end{figure}
\begin{figure}[h]
\includegraphics[width=\linewidth]{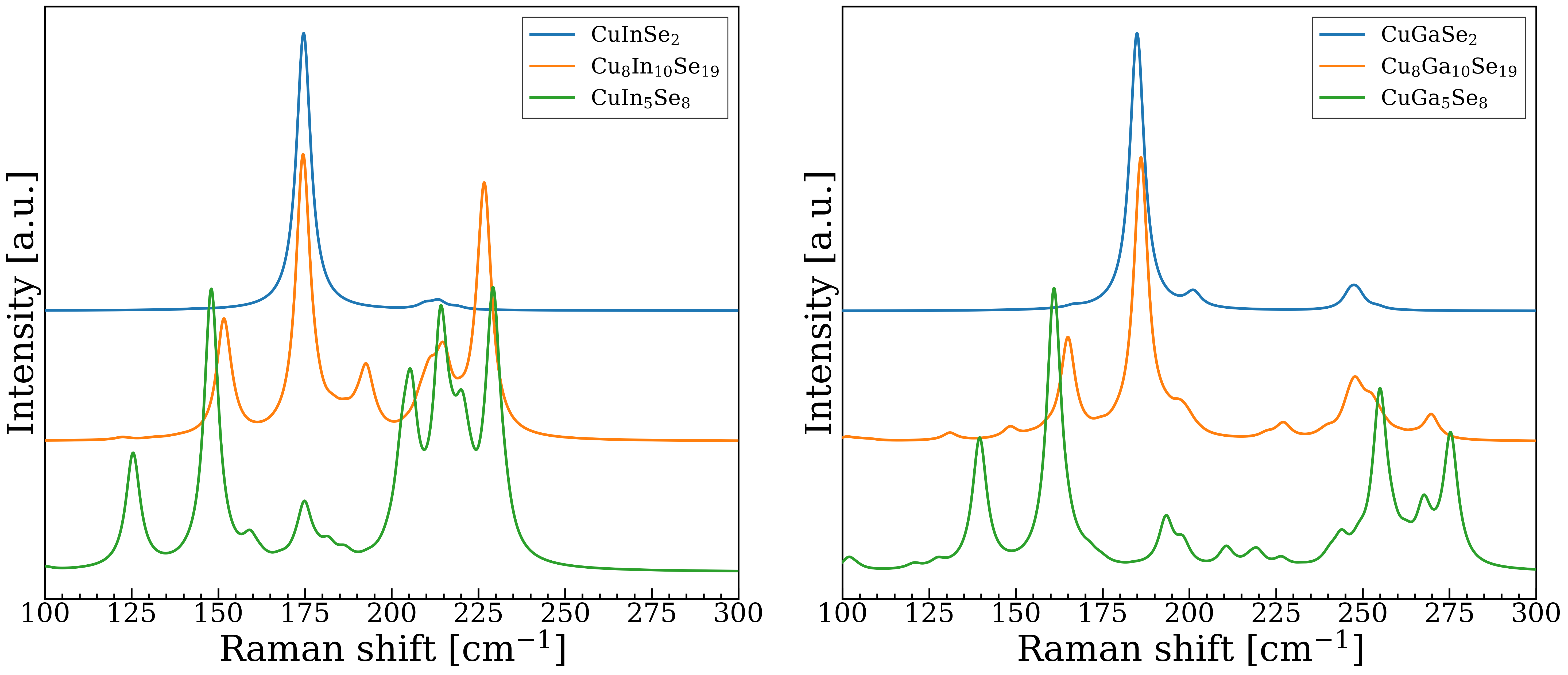}
\caption{Off-resonance Raman spectra simulated for three selected structures representing stoichiometric chalcopyrite CIGS (CuInSe$_2$ and CuGaSe$_2$), model OVC (CuIn$_5$Se$_8$ and CuGa$_5$Se$_8$), and off-stoichiometric CIGS with planar Cu-poor defects (Cu$_8$In$_{10}$Se$_{19}$ and Cu$_8$Ga$_{10}$Se$_{19}$). The simulations were performed using the \url{vasp_raman.py} utility developed by Fonari and Stauffer \cite{Stauffer2013}. For the input, the phononic frequencies and the macroscopic dielectric tensors were computed according to the density functional perturbation theory \cite{Baroni2001}, whereas the derivatives of the dielectric tensor were obtained numerically by displacing the atoms by 0.01 Å. All these calculations were carried out using the PBE+U functional (with the Hubbard U correction of 5 eV applied on Cu $3d$ electrons), after relaxing ionic positions (under the force threshold of 1 meV/atom), while keeping the cell parameters fixed to the PBEsol-computed values. The simulated unit cells contained 16, 28, and 37 atoms in the stoichiometric chalcopyrite, model OVC, and off-stoichiometric chalcopyrite models, respectively. To replicate the experimental peak broadening, the output Raman modes were smeared using the Lorentz function with the full width at half maximum (FWHM) of 5 cm$^{-1}$. \label{Raman_theory}}
\end{figure}

Figure~\ref{Raman} shows the Raman spectrum of the CuIn$_{x}$Ga$_{(1-x)}$Se$_2$ sample employed in the ATAS experiments. The spectrum shows two prominent features, a shoulder at frequency 159(4)~cm$^{-1}$ and a narrow peak at 176.2(7)~cm$^{-1}$. The narrow peak at approximately 176~cm$^{-1}$ is in good agreement with previous Raman measurements of the optical A$_{1g}$ phonon mode of chaclopyrite phase, while the shoulder at 159~cm$^{-1}$ is consistent with those observed in ordered vacancy compounds (OVC)\cite{SHEU2016}. The relative amplitudes were fitted using the fit function,
\begin{gather}
f(\nu) = \sum_{j=1,2}\frac{\alpha_j}{\sqrt{2\pi}\sigma_j}e^{-\frac{(\nu - \nu_j)^2}{2\sigma_j^2}} +d,
\end{gather}
as $\alpha^\text{CIGS}_{A_{1g}} = 5\cdot 10^4 \pm 8.8\cdot 10^3$ a.u. and $\alpha^\text{OVC}_{A_{1g}} = 2.707\cdot10^4 \pm 9.4\cdot 10^3$ a.u.. In Fig. \ref{Raman_theory} simulated off-resonance Raman spectra are shown for the chalcopyrite and model OVC structures. For the calculated structures we recover the $A_{1g}$ of the chalcopyrite and model OVC close to the experimentally observed frequencies.

\subsection{Fit of electronic decoherence time $T_2^e$ and the scattering phase $\phi$}

From the lineout of the CB spectral region in Fig. \ref{fig:55}(a), the electronic decoherence time and phase were fitted using Eq. \eqref{fnaff2}. To fit the decoherence time,  a bandpass filter was applied to remove the slow varying component of the $\Delta A$ originating from the coherent phonon motion. The band pass filter also reduces high frequency noise. Subsequently, for the temporal intervals 9.9-112.2 fs, 211.2-310.2 fs, 412.5-511.5 fs, the oscillation amplitude was fit using a least squares fit with fit function $f(\tau) = a\cos(\omega\tau +\phi)$. This produces the oscillation amplitudes in Fig. \ref{fig:55}(c), from which the decoherence time was fitted with an exponential decay using a weighted least squares fit where the uncertainties of the oscillation amplitudes were used as weights. In addition, for the temporal interval, 9.9-112.2 fs, the fitted phase $\phi$ of 4.1(3) rad was used to calculate the time delay of $t_W = 12.1(8)$ fs.

\subsection{Fourier map of scan in Fig \ref{fig:55}(b)}

\begin{figure}[htb!]
\includegraphics[width=0.5\linewidth]{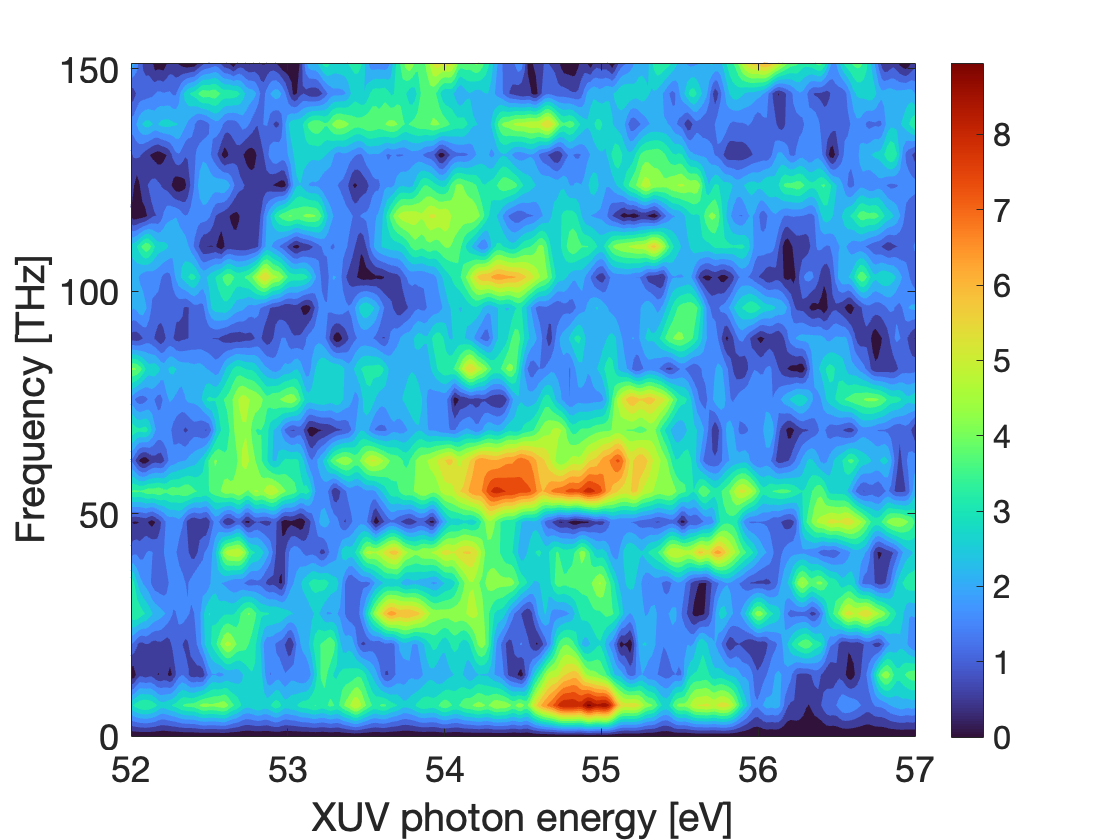}
\caption{Amplitude of Fourier transform of ATAS scan shown in Fig. \ref{fig:55}(b). The DC component was subtracted for visibility. The Fourier amplitude of frequency 55 THz has an onset for XUV photon energies greater than 54 eV.}\label{fig:fouriermap}
\end{figure}

In Fig. \ref{fig:fouriermap}, the Fourier transform over pump-probe delay is shown for the scan in Fig. \ref{fig:55}(b). A clear Fourier amplitude is shown around 55 THz for energies $>$ 54 eV. We therefore attribute these oscillations to originate from core level transitions to the CB. In Fig. \ref{fig:fouriermap}, The DC component was subtracted for visibility.

\subsection{XUV absorption spectrum of CIGS}

\begin{figure}[htb!]
\includegraphics[width=0.5\linewidth]{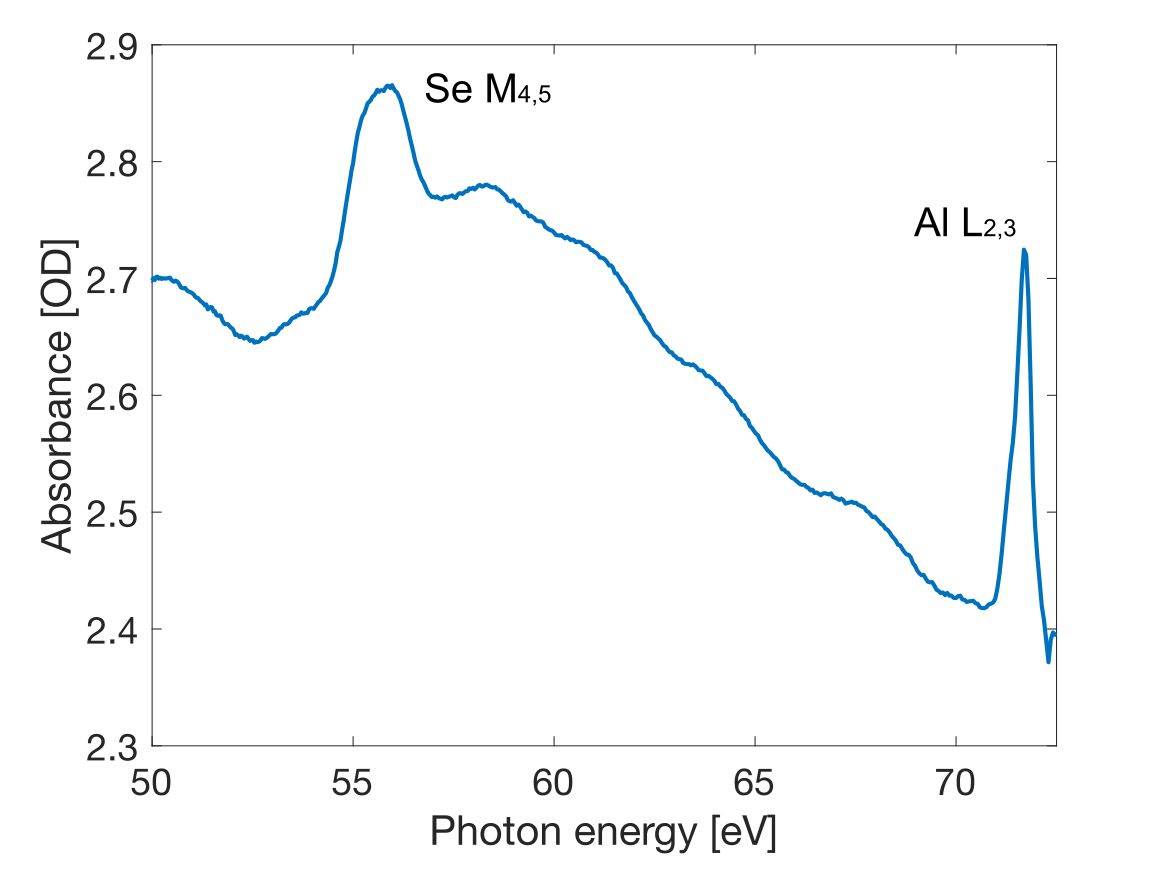}
\caption{XUV Absorbtion spectrum of CIGS measured as a function of XUV photon energy. The spectrum shows a broad absorption peak around 55 eV, with distinct features corresponding to the transitions from the $3d_{5/2,3/2}$ core levels to the conduction band. The Al L$_{2,3}$ edge is visible around 71.7 eV \label{fig:7}}
\end{figure}

In Fig. \ref{fig:7} the XUV absorption spectrum of the CIGS sample is shown. The Se $M_{4,5}$ absorption edge is visible around 55 eV with two distinct peaks corresponding to the XUV transitions from the $3d_{5/2,3/2}$ core to the conduction band.  In addition, the Al L$_{2,3}$ is measured at 71.7 eV. This edge originates from the 10 nm Al$_2$O$_3$ layer added on top of the CIGS layer to prevent the CIGS from oxidizing. The Al$_2$O$_3$ edge exhibited no transient response, due to it being an insulator.

\subsection{X-ray diffraction characterization}

To verify that the CIGS morphology was not transformed by the intense IR pulses, XRD measurements were performed after the ATAS measurements. 

\begin{figure}[h]
\includegraphics[width=0.5\linewidth]{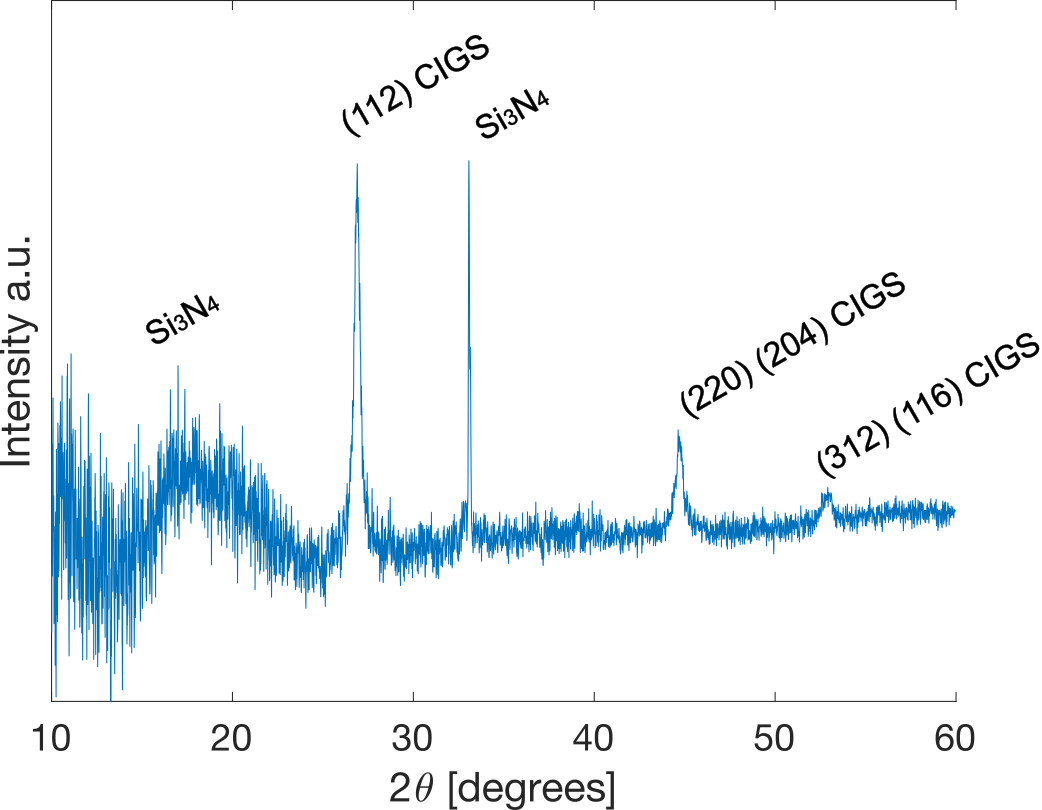}
\caption{XRD measurements on the CIGS sample studied in this work. \label{fig:xrd}}
\end{figure}

In Fig. \ref{fig:xrd}, the XRD measurement is shown. The XRD peaks are compared to previous XRD measurements of CIGS thin films, showing excellent agreement\cite{Dhage2011}. We therefore conclude that the samples were of high quality and the morphology was not changed during the scope of the measurement. We identify two main contributions from the Si$_3$N$_4$ substrate, a broad peak at angles ($2\theta$) in the range 15 to 25 degrees and a narrow peak at 33.07 degrees. We identify peaks corresponding to the CIGS chalcopyrite phase at angles 26.9, 44.64 and 52.91 corresponding to the Miller indices (112), (220)/(204) and (312)/116. The XRD characterization hence confirms that the CIGS sample did not transform from the chaclopyrite phase during the scope of our measurements.

\subsection{Band structure series and calculation of conduction band offset}
To investigate the dependence of the conduction band offset on the [Cu]/([Ga]+[In]) ratio, the Cu-Ga-Se$_2$ and Cu-In-Se$_2$ band structures were calculated for a set of structures with varying Cu to In/Ga concentration. In Fig. \ref{fig:bandstructures}, the band structures for the, 1:1:2, 10:12:23, 8:10:19, 6:8:15 and 4:6:11 structures are shown. As the Cu concentration decreases, a saddle point valley emerges at the Z-point with energy minima decreasing with decreasing [Cu]/([Ga]+[In]) ratio.
\begin{figure}[htp]
\includegraphics[width=0.8\linewidth]{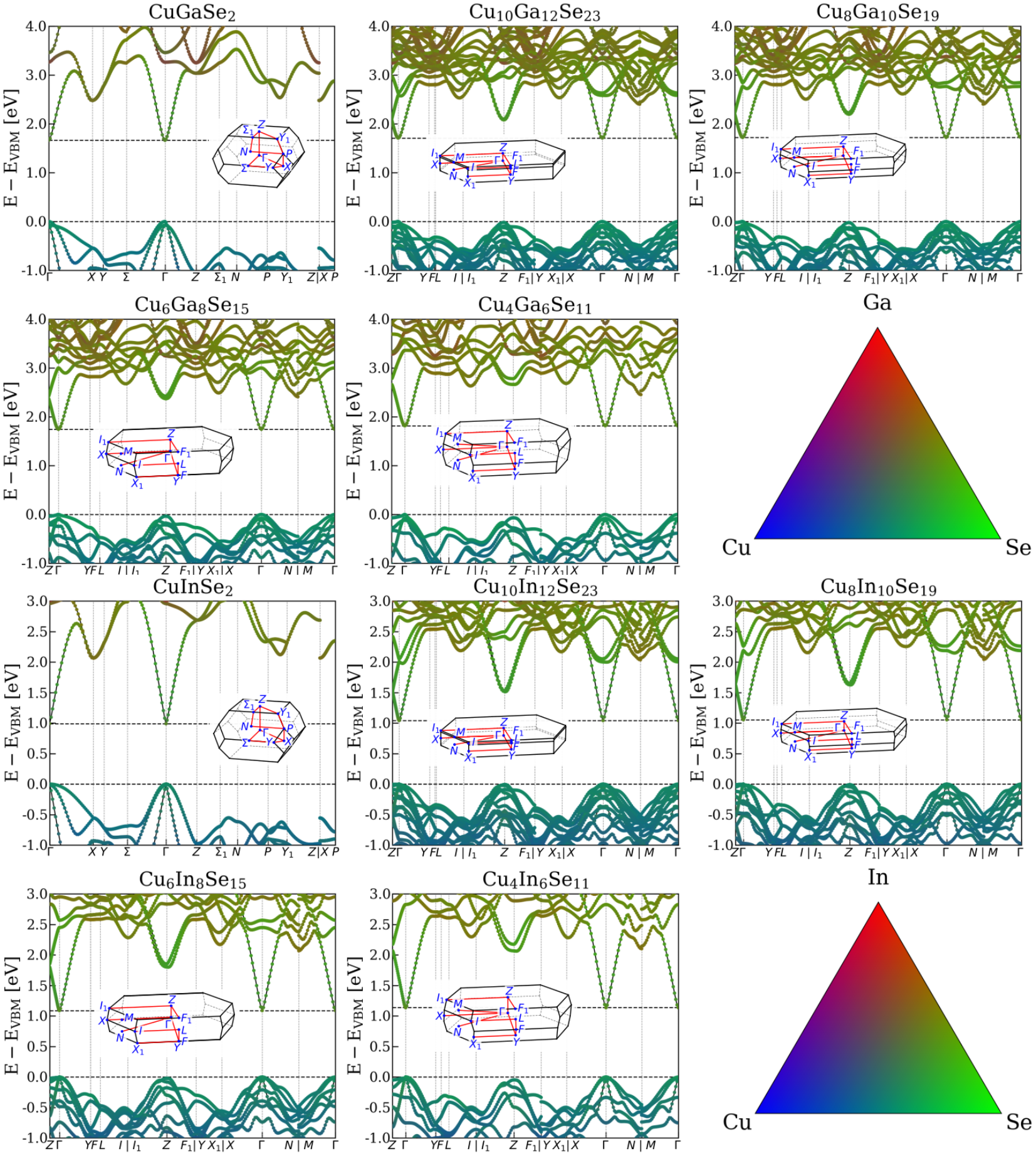}
\caption{Element-projected band structures for a series of (a) Cu-In-Se and (b) Cu-Ga-Se structures with different [Cu]/([Ga]+[In]) ratios. The structures are produced by varying the separation between the Cu-poor planar defects depicted in Figure \ref{fig1}. The insets illustrate the k-paths in the Brillouin zones. The band structures were computed using the PBE+U functional (with the Hubbard U correction of 5 eV applied on Cu 3d electrons), using the structures (both lattice parameters and ionic positions) optimised using the PBEsol functional. The band gaps were additionally adjusted to the values computed using the HSE06 functional, upshifted by 0.17 and 0.27 eV for the Cu-In-Se and Cu-Ga-Se structures, respectively.}
\end{figure}
In Fig. \ref{fig:dCBM} the calculated energy difference of the band energy minima of the Z and $\Gamma$ points are shown for Cu-In-Se (blue points) and Cu-Ga-Se (red points). As the Cu concentration decreases the conduction band minimum offset decreases, and by extrapolating the value (solid lines) to the Cu concentration of the CIGS sample in this study ([Cu]/([Ga]+[In] = 0.9) the predicted conduction band offset is retrieved as $\Delta E^{th}_{CB} = 0.25$ eV (denoted with a star).
\begin{figure}[htp]
\includegraphics[width=0.5\linewidth]{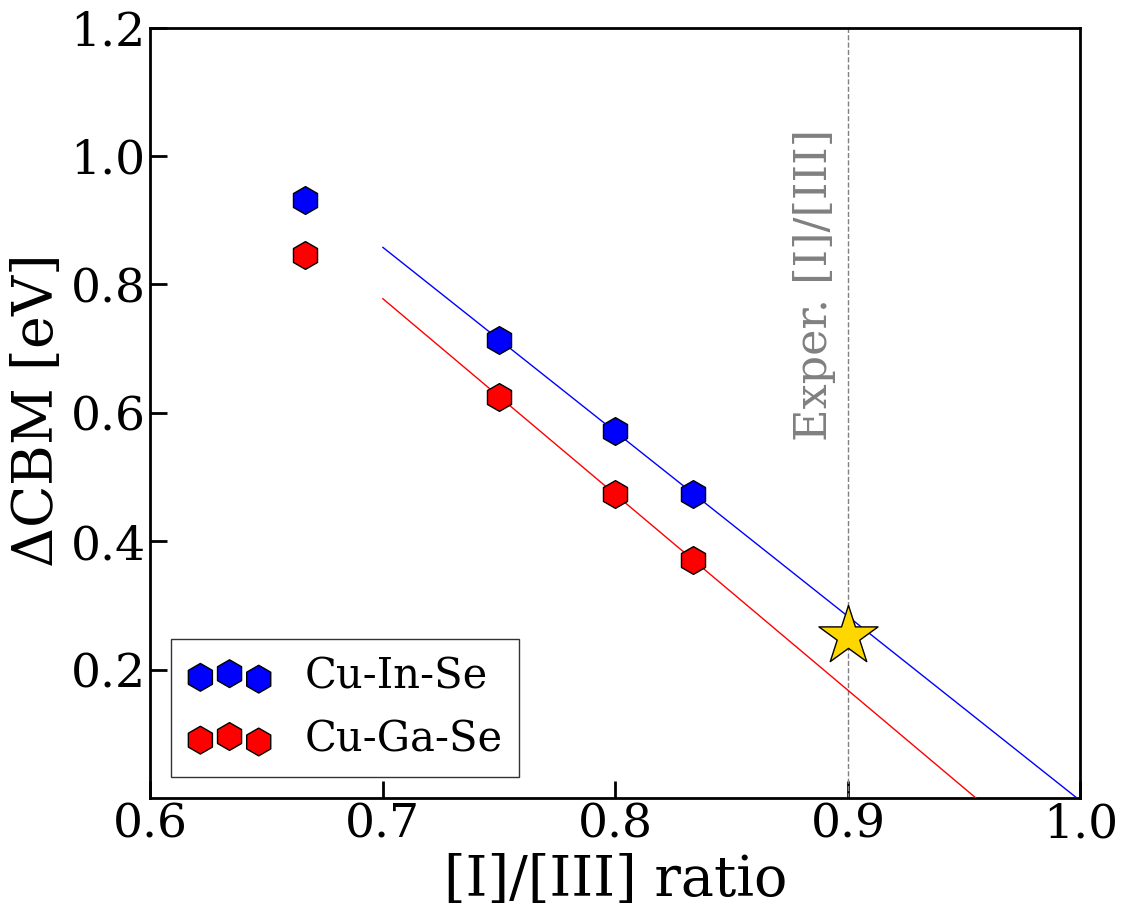}
\caption{Computed energy difference of the lowest unoccupied band at the high-symmetry Z and $\Gamma$ points for the Cu-In-Se and Cu-Ga-Se structures with different [I]/[III] $\equiv$ [Cu]/([Ga]+[In]) ratios. The hexagonal markers depict the values extracted from Figure \ref{fig:bandstructures}, and the lines represent linear fits for the three values closest to the ideal chalcopyrite stoichiometry. The yellow star marker at $\Delta E^{th}_{CB}=0.25$ eV is an extrapolation for the CIGS film composition used in experiments. \label{fig:dCBM}}
\end{figure}

\end{document}